     \documentclass[12pt]{article}
\usepackage[utf8]{inputenc}
     \usepackage[margin=1in]{geometry}
     \usepackage{times}
     \usepackage{lipsum}
     \usepackage{ragged2e}
\usepackage{setspace}
\spacing{1.5}

\usepackage{sectsty}
\usepackage[italicdiff]{physics}
\usepackage{titling}

\usepackage{amsmath}

\usepackage [english]{babel}

\usepackage{array}
\usepackage{pdflscape}
\usepackage{booktabs}
\usepackage{rotating} 
\usepackage{graphicx} 

\usepackage{etex}
\usepackage{amsthm}
\usepackage{amsmath}
\usepackage{amssymb}
\usepackage{soul}
\usepackage{calc}
\usepackage[usenames,dvipsnames]{xcolor}
\usepackage{mathptmx}
\usepackage{colortbl}
\usepackage[boxed]{algorithm2e}
\usepackage{epstopdf}
\usepackage{arydshln}
\usepackage{booktabs}
\usepackage{natbib}
\usepackage{hyperref}

\hypersetup{
	colorlinks=true,         
	linkcolor=MidnightBlue,          
	citecolor=MidnightBlue,
	urlcolor=MidnightBlue            
 }

\usepackage{breakurl}
\usepackage{bookmark}
\usepackage{etoolbox}
\usepackage{graphicx}
\theoremstyle{plain}
\theoremstyle{definition}
\newtheorem{definition}{Definition}
\usepackage[title]{appendix}
\usepackage{mathtools}
\usepackage{stackengine}
\usepackage{cancel}
\usepackage{comment}

\newtheorem{thm}{Theorem}[section]

\usepackage{authblk}
\pretitle{
\centering

  \vspace{1.5cm}
  \LARGE
  }

\posttitle{%
\centering
  \par\vspace{0.4em}
  \LARGE\itshape
}
\preauthor{%
\centering
  \par\vspace{1.2cm}
  \normalsize
}
\postauthor{\par}
\centering
\predate{%
  \par\vspace{1em}
  \small\scshape
}
\postdate{%
\centering
  \vspace{1cm}

}

\title{\textit{Apples Falling, Buckets Rolling, and Why Inertia Keeps Trolling}:\\
       Inertial Motion is Not Natural Motion}

\author[1,2]{Nicola Bamonti}
\date{\textsc{\textbf{Provisional Draft}}\\Please, do not circulate without my consent}

\affil[1]{Department of Philosophy, Scuola Normale Superiore, Piazza dei Cavalieri, 7, Pisa, 56126, Italy}
\affil[2]{Department of Philosophy, University of Geneva, 5 rue de Candolle, 1211 Geneva 4, Switzerland}

\begin{document}

\maketitle


\begin{abstract}
\singlespacing

Inertia has long been treated as the paradigm of natural motion.
This paper challenges this identification through the lens of General Relativity.
Drawing on \cite{Norton2012}’s distinction between idealisation and approximation and analysing key results from \cite{Tamir2012} on the theorems of  Geroch–Jang, Ehlers–Geroch, Einstein–Grommer, and Geroch–Traschen, I argue that geodesic motion—commonly treated as the relativistic expression of inertia—fails to qualify as either.
Rather, geodesic motion is best understood as a \textit{useful construct}—a formal artefact of the theory's geometric structure, without real or fictitious instantiation, and excluded by the dynamical structure of GR.
In place of the inertial motion, I develop a layered account of \textit{natural motion}, which is not encoded in a \textit{single} \lq{}master equation of motion\rq{}. Extended, structured, and backreacting bodies require successively refined dynamical formalisms that systematically depart from geodesic motion.
This pluralist framework displaces geodesic motion as the privileged expression of pure gravitational motion, replacing it with a dynamically grounded hierarchy of approximations fully consistent with the Einstein field equations.
Inertial motion thus emerges not as the universal default of motion under gravity alone, but as a formal construct that stands apart from the pluralistic framework in which natural motion is genuinely realised.
\end{abstract}

\tableofcontents

\clearpage

\setstretch{1.2}

\justifying
\section{Introduction}\label{intro}

It is 1666, and England is in the grip of the Great Plague. That year, Cambridge University was closed because of the epidemic, and Newton, then a young student, retired to his family estate at Woolsthorpe Manor. Beneath an apple tree, immersed in solitary thought, he contemplated the puzzles of optics and infinitesimal calculus.
An apple falled.
\textsl{Eureka!} -- he exclaimed.
\textsl{The force that draws the apple earthward is the same as that which holds the Moon in its orbit}, Newton thought.
\textsl{And that's not all} --- he continued --- \textsl{the tendency of a body to resist changes in its state of motion explains why the surface of water in a rotating bucket becomes concave when the bucket rotates with respect to absolute space.}
In that moment, the qualitative insights of Galileo and Descartes crystallised in rigorous form in his mind, and the three laws of dynamics manifested, including the first: the law (or principle) of inertia.

This, of course, is just a tale rather than a literal report of historical events.
Newton stands in continuity with a long tradition of thinkers in the West, known as \textit{natural philosophers}, who questioned Nature and its laws.
Among these questions is that pertaining to the motion of bodies: a topic that have occupied natural philosophers for over two millennia.

Beyond the descriptive query—\textit{how} do bodies move?—lies a perhaps more profound concern: \textit{why} do bodies move as they do? What is the nature of their motion?

Notably, these questions were often posed in terms of a dichotomy between motions caused by agents external to the bodies and motions that arise intrinsically when no external cause acts.
Reformulated, these inquiries becomes: \textit{how} do bodies behave in the absence of external influences, and what governs their motion when such influences are present?
And \textit{why} do they assume such and such a configuration of motion?

Aristotle was the first to investigate what we now call inertia with his distinction between \textit{natural motions}, characteristic of bodies \lq{}left to themselves\rq{}, and \textit{violent motions}, which result from external influences \citep{sachs1995aristotle}.
In the late Middle Ages, John Buridan advanced the \textit{impetus theory}, positing that a body in motion carries an internal \textit{impetus} that sustains its motion even in the absence of external causes \citep{Jung2011}. 
This marks historically the first known articulation of a concept recognisably akin to inertia, later refined by Galileo and Descartes, and rigorously formalised by Newton in his \lq{}first law\rq{}.\footnote{For an analysis on the origin and role of the Law of Inertia in Newton's thinking see \cite{Earman1973}.}

In the twentieth century, Einstein extended the trajectory begun by Galileo and Newton. 
Within General Relativity (GR), he introduced the \textit{geodesic principle} as a relativistic analogue of Newton’s first law: free bodies move along geodesics of a curved spacetime.
GR thereby seemed to elegantly answer not only the question of \textit{how} free bodies move, but also—at least to some degree—\textit{why} they do so \citep{Weatherall2011,Weatherall2016-WEAIME}.\footnote{The question of whether—and in what sense—GR \textit{explains} inertial motion has received renewed attention in recent literature, beginning with \cite{Brown2005-kq}. Brown does not commit to any specific philosophical theory of explanation; rather, he adopts a liberal usage in which \lq{}explanation\rq{} may involve offering \textit{sufficient} conditions for a phenomenon, providing conceptual \textit{insight} into it, or showing how it can be \textit{formally derived}. For a broader overview of philosophical accounts of scientific explanation, see \cite{sep-scientific-explanation}.
\cite{Weatherall2016-WEAIME} advocates the so-called \textit{Puzzle-ball conjecture}, which proposes that the explanatory structure of GR should not be understood in terms of asymmetric derivation from more fundamental axioms. Instead, explanation is viewed in terms of \textit{interconnectedness} among the theory’s core principles. On this view, EFEs and the geodesic principle are interdependent: GR explains geodesic motion by appeal to Einstein’s equations, but one could equally argue that Einstein’s equations are explained by the geodesic principle. As Weatherall puts it: \lq{}\lq{}[G]eneral relativity explains inertial [i.e. geodesic] motion by appeal to Einstein’s equation, but it may equally well explain Einstein’s equation by appeal to the geodesic principle\rq{}\rq{} (ibid., p.38).}

This paper revisits the concept of inertia through the lens of GR, tracing its development from Galileo to Einstein, with a particular focus on the status of inertial motion in the relativistic framework.
I argue that inertia is best understood not as a universal feature of natural motion, but as a \textit{useful construct}: a formal artefact that masks the more complex dynamical regimes governing the motion of free bodies.

To make this case, I adopt the methodological distinction introduced by \cite{Norton2012} between \textit{idealisation} and \textit{approximation} (see also \citealp{Bamonti2023}). 
An approximation describes a real target system inexactly. An idealisation, by contrast, refers to a \textit{different, new} system—often fictitious—whose exact properties resemble some aspects of the target system. Both require referents: approximations presuppose a real target system; idealisations presuppose a (real or) fictitious new one. The crucial difference between the two concepts is that only idealisation introduces a \textit{new} system; approximation simply misdescribes the existing target.

With this distinction in hand, I argue that geodesic motion in GR qualifies as neither. 
It is not an approximation, because it fails to inexactly describe the behaviour of any admissible system within spacetime. 
Nor is it an idealisation, since there is no consistent limit system—real or fictitious—within the theory that bears the geodesic property.\footnote{Frequently throughout the rest of this paper, with a certain abuse of terminology, I will say that \lq{}geodesic motion is not an allowed idealisation\rq{}, rather than saying that \lq{}a body that follows geodesic motion is not an acceptable idealisation\rq{}, effectively attributing the idealisation to the property of motion and not to the system that instantiates that property.} 
As I will show, geodesic motion is defined \textit{off-shell}: it is not derived from the dynamics of any solution to the EFEs. Moreover, the associated trajectories lie \textit{outside the manifold}.
In both respects, geodesic motion lacks an instantiating referent. 
This disqualifies it not only as an approximation (which requires a real target system), but also as an idealisation (which requires a coherent surrogate system that bears the relevant property). 
The failure is twofold: either the system violates the theory’s dynamics, or it is not a system at all.

To support this claim, I draw selectively on \cite{Tamir2012}’s analysis of attempts to derive the geodesic principle. 
Among the many strategies he surveys, I focus on four paradigmatic cases:  the \textit{Geroch–Jang theorem}, the \textit{Ehlers-Geroch theorem}, the \textit{Einstein–Grommer derivation}  and the \textit{Geroch–Traschen theorem}.
Taken together, these cases demonstrate that geodesic motion is  neither the limiting behaviour of any real system governed by GR, nor a property of any coherent idealised system within its scope.

Accordingly, I propose that geodesic motion be reclassified as a \textit{third category}: a \textit{useful construct}—a geometrically defined property intrinsic to the theory’s formalism, indispensable for many theoretical purposes, but not instantiated—nor instantiable—by any physically meaningful model, \textit{whether real or fictitious}. This third category, which lies outside Norton’s original dichotomy, helps clarify why geodesic motion, despite its formal utility, fails to describe, approximate, or idealise the natural motion of free bodies.

By contrast, I argue that \textit{natural motion} should be understood as a \textit{layered concept}: a hierarchy of increasingly refined dynamical regimes, grounded in physically admissible solutions to Einstein’s equations. Depending on their internal structure, spatial extension, and gravitational self-interaction, bodies require distinct formalisms.
In the simplest case, a spinning test body experiences deviations from geodesic motion, reflecting sensitivity to curvature gradients governed by the Mathisson–Papapetrou–Dixon (MPD) equations  (§\ref{MPD}).
These effects are complemented by the geodesic deviation equation, which models tidal forces across a congruence of worldlines composing the body (§\ref{deviation}). 
When backreaction is included, further deviations arise: first in the perturbative regime—via formalisms like MiSaTaQuWa (§\ref{perturbed})—and ultimately in the nonlinear regime, where the body's stress-energy sources the spacetime metric.

These are not successive refinements of geodesic motion, but systematic \textit{replacements}. Unlike the geodesic principle, they describe bodies that exist within spacetime and respect the theory’s dynamical constraints.

Natural motion is not inertial motion \lq{}de-approximated\rq{} or \lq{}de-idealised\rq{}, but a fundamentally plural framework, stratified by structural complexity and dynamical interaction. Each level in this hierarchy defines a physically grounded regime of approximation. Geodesic motion does not appear at any level of this hierarchy. It stands apart from it.

This interpretive shift preserves the representational power of GR while reframing the geodesic principle itself: not as a fundamental principle of motion, but as a mathematical artefact—elegant and illuminating, but ultimately unrealised, and belonging to the theory’s formalism rather than its ontology.

\paragraph*{Roadmap.} The paper proceeds in three main stages.

In \S \ref{classicalinertia}–\ref{geodesic}, I examine classical and relativistic formulations of the Principle of Inertia, showing that they collapse into circularity or triviality.  

In \S\ref{THEOREMS},  I assess whether the geodesic principle can be derived from within GR itself. I begin with the Geroch–Jang theorem, which shows that geodesic motion can be assigned to certain matter distributions under highly restricted assumptions. As Tamir argues, however, this result presupposes a test-body regime that is not dynamically justified within the field equations.
I then turn to the Ehlers–Geroch theorem, which constructs a well-defined limit system by considering a sequence of spacetimes whose matter content becomes increasingly concentrated. Yet in the limit, the matter content vanishes or the field equations are violated.
In the Einstein–Grommer strategy, geodesic motion is attributed to a singularity excised from the manifold. Since the trajectory lies outside spacetime, there exists no real system that it could approximate.
The Geroch–Traschen theorem complements this verdict, demonstrating that the limit system associated with a massive point particle does not belong to the space of admissible solutions to Einstein’s equations.

In \S\ref{EXTENSION} and \ref{REALISTIC}, I turn to the natural motion of free bodies, albeit approximate. In §\ref{EXTENSION}, I consider structured, spatially extended but non-backreacting systems—true test bodies—and show, using the MPD formalism and geodesic deviation, that geodesic motion already fails in this regime.
In §\ref{REALISTIC}, I examine backreacting bodies. Perturbative treatments such as MiSaTaQuWa formalism capture self-interaction, while the cosmological case of FLRW dust—though fully non-linear—illustrates the limits of geodesic motion in backreacting regimes.

Finally, in \S\ref{naturalmotion}  I synthesise these layers into a unified interpretative framework. Natural motion is a stratified concept: a sequence of approximated dynamical regimes, each valid for a class of bodies with specific structural features. Geodesic motion is not the base layer of this hierarchy. It is excluded from it.

\section{Inertial Motion and the Principle of Inertia: The Classical Story}\label{classicalinertia}

The concept of inertial motion has long served as a cornerstone of classical mechanics and continues to hold foundational significance in modern physics. A widely cited formulation, often retrospectively associated with Galileo, states:

\begin{definition}
    \textbf{Inertial Motion (Version 1)}: A body undergoes inertial motion if and only if it is in uniform motion and continues to move uniformly, or it is at rest and continues to remain at rest. \label{INERTIA1}
\end{definition}

Although Galileo never articulated a formal \textit{definition} of inertia in the modern sense, in \textit{Two New Sciences} (\citeyear{galilei1914dialogues}), he argued—through thought experiments and observations (notably involving inclined planes)—that in the absence of resistance, a moving body would continue to move at constant speed in a straight line. These insights laid the foundation for what Newton would later formalise as the Principle of Inertia (\textbf{PIN}) in the \textit{Principia} (\citeyear{newton1687principia}), which seeks to capture the \textit{regularity} of how bodies move when not subject to external influences. This principle has been formulated in two distinct ways:
\begin{definition}
    \textbf{PIN (v.1)}: Bodies maintain inertial motion if and only if no external \textit{force} acts upon them. \label{PIN1}
    \end{definition}   \begin{definition}
    \textbf{PIN (v.2)}: Bodies \textit{sufficiently} distant from other bodies retain their state of inertial motion \citep{einstein2015relativity}. \label{PIN2}
\end{definition}

Both versions, however, face conceptual difficulties that compromise the foundational clarity of the notion of inertia.

In Definition \eqref{PIN1}, the term \lq{}\textit{force}\rq{} is itself defined via deviation from inertial motion—making the formulation circular: \lq{}\lq{}\textit{Bodies maintain inertial motion if they do not deviate from inertial motion}\rq{}\rq{}. 
Similarly, Definition \eqref{PIN2} invokes \textit{sufficient distance}, but sufficiency here is implicitly defined as  \lq{}enough to move inertially\rq{}, rendering the definition equivalent to: \lq{}\lq{}Bodies which move inertially retain their state of inertial motion.\rq{}\rq{}

A natural refinement of these definitions introduces the notion of \textit{inertial reference frames}. This requires first specifying what constitutes a reference frame, and then what qualifies as \textit{inertial}.

Adopting an operational perspective following \cite{Bamonti2023}, a \textit{(spatiotemporal) reference frame} may be defined as a set of four degrees of freedom $\{x^I\}_{I=1,\dots,4}$ \textit{instantiated} by a physical system (typically three rods and one clock), which yields a local diffeomorphism  $U\subseteq \mathcal{M} \to \mathbb{R}^4$, where $U \subseteq\mathcal{M}$ is an open region of a differentiable manifold $\mathcal{M}$.
Each point $p\in U$ is thereby \textit{uniquely} assigned four real numbers.

This definition does not require necessarily the domain of the diffeomorphism to be a subset of the bare manifold alone.
The region $U$ can also be equipped with additional geometric structure appropriate to the physical theory under consideration. 
Therefore, depending on the theory considered, a reference frame could be understood as a map whose domain is a \textit{structured space}.

For instance, in theories with fixed spatiotemporal background, such as special relativistic ones, a reference frame acts as a \textit{global Poincaré} map that preserves the structure $(\mathcal{M},\eta_{ab})$, where  $\eta_{ab}$ denotes the flat Lorentzian metric:
\begin{equation}
x^I:(\mathcal{M},\eta_{ab})\to\mathbb{R}^4.
\end{equation}
Analogously, in covariant Newtonian theory, the reference frame acts as a \textit{global Galilean} map:\footnote{$\xrightarrow{\sim}$ denotes an isomorphic mapping between spaces.}
\begin{equation}
x^I:(\mathcal{M},t_{ab},h^{ab},\nabla)\xrightarrow{\sim}\mathbb{R}\times\mathbb{R}^3\to\mathbb{R}^4.
\end{equation}

Here, spacetime is equipped with classical structure $(t_{ab},h_{ab},\nabla)$: defining a classical spacetime with absolute time (encoded via the degenerate temporal metric $t_{ab}$), absolute space (encoded via the spatial metric $h_{ab}$), and a \textit{compatible} flat, torsion-free connection $\nabla$ satisfying $\nabla_a t_{bc}=0; \nabla_ah_{bc}=0$.\footnote{$t_{ab}$ has signature $(1,0,0,0)$ and $h^{ab}$ has signature $(0,1,1,1)$. The flat connection $\nabla$  satisfies \ $R^{a}_{\;bcd}=0$ and is one of the infinitely many \textit{compatible} flat connections.} 

In all such cases, the reference frame must preserve the symmetries of the geometric structure: Poincaré symmetry in special relativity, Galilean symmetry in Newtonian mechanics. It must also be adapted to the flat affine structure that, in the classical setting, provides the background for defining inertial motion.\footnote{In formal terms, any reference frame is physically significant only insofar as it preserves the automorphisms of the structured spacetime.In Newtonian theory, for example, only inertial frames preserve \textit{Galilean symmetries}, which are also dynamical symmetries of the theory, as per Earman's SP principles \citep{Earman1992-jn}. As \cite{GomesSymmetries} notes: \lq{}\lq{}reference frames, spacetime symmetries and dynamical symmetries are given together as a package-deal\rq{}\rq{}. \label{packagedeal}}

Given this structured background, an inertial reference frame may now be defined as follows:\footnote{The \textbf{INRF} $\{x^{I}\}$ may also be considered \lq{}attached\rq{}to the body undergoing inertial motion. This point will be relevant in \S\ref{geodesic}, where inertial frames in GR are shown to be only \textit{locally} defined and necessarily \textit{comoving} with the body.}

\begin{definition}
    \textbf{INRF (v.1)}: An inertial reference frame provides a standard of space and time measurement relative to which:
\begin{itemize}
\item[(i)] Bodies not subject to \textit{net external forces} move uniformly $\bigl( \frac{dx^{i=1,2,3}}{dt}=\text{const.}$, with $t\equiv x^4\bigr)$;\footnote{In Newtonian framework, time is absolute and global. Hence the temporal parameter is the same in all inertial frames.} 
\item[(ii)] Accelerated motion obeys Newton's second law: $F=m\frac{d^2x^i}{dt^2}$;
\item[(iii)] No fictitious forces (e.g., centrifugal, Coriolis) are present.
\end{itemize} \label{INRF}
\end{definition}

On this basis, one can restate the principle of inertia as follows:

\begin{definition}
    \textbf{PIN. (v.3)}:  \textit{Relative to} an inertial reference frame (as defined above in Def. \eqref{INRF}), bodies either maintain inertial motion (uniform velocity or rest) or accelerate in accordance with Newton's second law. \label{PIN3}
\end{definition}

However, this reformulation still hinges on the very notion it purports to clarify, namely, inertia, thereby reintroducing the original circularity. 
The definition of an inertial frame appeals to the behaviour of \textit{unforced} bodies, but what counts as unforced depends once again on whether the motion is inertial.  If we already know which bodies are unforced, we already know which bodies are inertial.
A more fundamental reconsideration of both inertial motion and inertial frames is therefore required.

\subsection{Principle or Law?}\label{lawvsprin}

Inertia has led a curious double life in the history of physics. On one hand, it has been treated as an empirical \textit{law} describing observed regularities in motion; on the other, as a \textit{principle} that accounts for or constrains those regularities.
This duality warrants scrutiny. To treat inertia as a law of motion is to misrepresent its foundational role; yet to regard it as a purely a priori principle, immune to empirical revision, is to obscure its empirical origin.

Thus far, I have referred to the \lq{}principle of inertia\rq{} rather than the\lq{}law of inertia\rq{}.
It is now necessary to clarify this distinction. 
 I introduce two interpretive modes: what I call a \textit{law-like terminology}, which treats inertia as a descriptive regularity (whether grounded in forces or geometry), and a \textit{principle-like terminology}, which treats inertia as a constitutive constraint on admissible dynamics.

\paragraph{The Law of Inertia: Law-like Terminology.}

In Newtonian mechanics the \lq{}law of inertia\rq{} is traditionally articulated as \textit{Newton’s first law}: that free (i.e. force‑free) bodies continue in uniform straight‑line motion (cf. Def. \eqref{PIN1}).

This formulation belongs to the standard \textit{three-dimensional framework}, in which inertial frames are those relative to which Newton’s laws take their simplest form: free particles exhibit uniform motion  $(v=\rm const)$, while forced particles obey $F=ma$. As \cite{Weatherall2020-WEATDO-7} notes, however, the appeal to \textit{simplicity} as a criterion for frame selection is methodologically ambiguous.

This setting falls within the category that I call the \textit{law-like interpretation} of inertia: the idea that inertia is a contingent, descriptive law.

Within this broad category, \cite{JacobsForthcoming-JACHNT-2} identifies a more specific \textit{law-based approach}, in which Newton's first law serves to \textit{define} inertial frames. This approach is found also in \cite{sep-spacetime-iframes}, where an inertial frame is defined as a spatial reference frame together with some means of measuring time, so that uniform motion can be distinguished from accelerated motion. In this view, an observer in an inertial frame sees \textit{non-inertial} bodies moving according to $F=ma$.
\lq{}\lq{}[\dots] An inertial frame is a reference-frame with a time-scale, relative to which the motion of a body not subject to forces is always rectilinear and uniform, accelerations are always proportional to and in the direction of applied forces, and applied forces are always met with equal and opposite reactions [\dots] in accord with Newton’s laws of motion. (ibid., p.1)\rq{}\rq{}.

This view aims to understand inertia as an empirical law, on a par with Newton’s Second and Third Laws. It \textit{describes} the motion of free bodies as uniform and rectilinear.\footnote{Whether the \textit{Law of inertia} constitutes an independent law or a corollary of Newton’s second law depends on the formulation of Newtonian mechanics.
In the four-dimensional covariant framework, once inertial frames are defined via the flat affine structure, the First Law follows trivially from the Second: force-free particles follow geodesics (i.e. \lq{}straight lines\rq{} in space–time).
By contrast, in the traditional three-dimensional formulation, \lq{}\lq{}the first law asserts that there exist certain frames with respect to which the second law is supposed to hold. The second law thus does not even make sense without the first law to define those frames\rq{}\rq{} \cite[p.7]{JacobsForthcoming-JACHNT-2}.}

However, Jacob's law-based approach is problematic.  
First, as also Jacobs argues, it risks \textit{circularity}: inertial frames are identified by the very behaviour (inertial motion) that the law is supposed to explain. This undermines the status of Newton’s first law as an empirical law on par with the others.\footnote{This problem did not afflict Newton himself, who formulated the first law relative to \textit{Absolute Space}, thereby circumventing any need for a dynamical definition of inertial frames.}
Second, since frame transformations can always be chosen to recover formal simplicity, the definition of inertial frames becomes overly liberal and lacks theoretical discipline.

An alternative, also analysed and ultimately rejected by Jacobs, is thee \textit{structure‑based approach}, which operates within a \textit{four-dimensional framework}.
Here, inertial frames are not defined by particle dynamics, but by their adaptation to a fixed background structure—specifically, Galilean spacetime equipped with a flat affine connection, an absolute time function, and a degenerate spatial metric.
A reference frame is said to be \textit{inertial} if it is adapted to this structure, in the sense that the affine connection has vanishing coefficients and the temporal and spatial metrics take their standard Pythagorean form. In such frames, free bodies follow geodesics of the connection, and their motion appears as uniform and rectilinear.

On this view, the \textit{law} of inertia is thereby grounded not in particle motion, but in spacetime geometry: it does not \textit{define} inertial frames; rather, it \textit{presupposes} them. 
Inertial motion then simply coincides with the statement that free bodies follow geodesics of the flat affine structure (cf. \citealp{Earman1973}). Newton’s Second Law then acquires its standard form, $F=ma$, in those frames that faithfully represent the background metric and in which force-free bodies move with constant coordinate velocity.

More precisely, the structure-based account proceeds as follows:
(i) It stipulatse a background spatiotemporal structure; 
(ii) It identifies certain frames that preserve this structure; 
(iii) It restricts the label \lq{}inertial\rq{} to those privileged frames, selected in (ii).

\textit{Nota Bene}: despite this geometric emphasis, the structure-based approach still belongs to the \textit{law-like} interpretation: it seeks to articulate a \textit{law} of inertia, albeit one grounded in geometry rather than particle dynamics.  It is important not to conflate Jacobs’ \textit{law-based} approach with the broader \textit{law-like} mode of interpretation I employ here.

\paragraph{The Principle of Inertia: Principle-like Terminology.}

A promising alternative to the law- or structure-based interpretations of inertia is offered by the \textit{symmetry-based} approach to inertia, also developed by \cite{JacobsForthcoming-JACHNT-2}.
Rather than grounding inertia in particle motion or background geometrical structures, this approach defines inertial frames as those that \lq{}\lq{}\textit{mesh with the dynamical symmetries of Newton theory} (ibid., p.2)\rq{}\rq. This marks a significant interpretive shift: inertia is no longer a descriptive law of motion, nor a geometrical consequence of a stipulated spacetime structure, but a constitutive feature of the theory’s dynamical symmetry group (cf. fn.\ref{packagedeal}).

On this account, an \textbf{INRF} is one in which Newton’s laws retain their \textit{form} under the relevant symmetry transformations, specifically, the Newton group, comprising time-independent translations, rotations, and boosts.

Although Jacobs refers to these as \textit{form-preserving} transformations, the underlying motivation is clearly solution-theoretic: the requirement is not merely that the laws \lq{}look the same\rq{} across frames, but that the Newton group relates entire families of dynamically admissible models. In this sense, Jacobs’ proposal aligns with the broader interpretation of symmetry as preserving \textit{solutionhood}—even if he does not frame it in those terms explicitly (cf. \citealp{BamontiGomes2024}).
Thus, the theory’s symmetries constrain both its dynamics and its class of admissible frames.

Jacobs’ formal definition is as follows:\footnote{Jacobs does not sharply distinguish between reference frames and coordinate systems—a conflation I avoid here. For a careful analysis of this distinction, see \cite{Bamonti2023}. I have accordingly reformulated his definition.}
\begin{quote}
    \textbf{Inertial [reference frame] (symmetry-based)}: a [reference frame] that is adapted to a symmetry-invariant metric, \textit{and in which force-free bodies move with constant velocity.} [my italics] (ibid., p. 22).
\end{quote}
Three clarificatory remarks are in order.

First, while Jacobs is explicit in identifying the circularity of the law-based approach—wherein one defines inertial frames via laws that themselves require inertiality—he does not extend this concern to his own symmetry-based formulation. Yet the potential problem remains: the notion of \textit{force} is itself defined as the cause of deviation from inertial motion. It cannot therefore figure in a non-circular definition of inertial frames. To invoke force in this context is to presuppose precisely what one is trying to define. Jacobs does not explicitly address this concern, but his derivation of the metric from symmetry constraints suggests a more principled basis for identifying the appropriate frames, thereby partially defusing the objection.

Second, Jacobs’ definition, while insightful, offers at most a \textit{necessary} condition for inertiality of reference frames, not a sufficient one.
An inertial reference frame that is \emph{dynamically uncoupled} from the system it describes may \textit{still} satisfy the symmetry-based definition, yet fail to \lq{}mesh\rq{} with the dynamical symmetries of the theory. That is because in case of uncoupled frames, spacetime and dynamical symmetries fail to coincide (see \citealp{BamontiGomes2024} for details).\footnote{Briefly, in frameworks where both the reference frame and the target system are modelled as dynamical fields, a frame is \textit{dynamically uncoupled} if one can apply a dynamical symmetry transformation to either component independently, while preserving solutionhood.}
As such, Jacobs’ proposal more accurately captures the notion of \emph{dynamical coupling} rather than inertiality proper.

Third, I maintain that Jacobs' symmetry-based definition naturally supports a \textit{principle-like} interpretation of inertia.
Unlike the law-based or structure-based accounts, which treat inertia as an empirical regularity or a geometrical fact,  the symmetry-based approach reframes inertia as a \textit{constitutive principle}: a constraint specifying which frames are admissible, and under what conditions the solutionhood of dynamical laws is preserved. 
The principle of inertia, thus understood, emerges not as a contingent fact but it constrains both the theory’s geometrical and dynamical structure.

Notice that Jacobs himself does not explicitly frame his proposal in these terms. He does not present his symmetry-based definition as a new principle of inertia, nor does he invoke the distinction between 'law-like' and 'principle-like' terminologies.
His aim is primarily diagnostic: to resolve the failure of standard accounts of inertial frames to distinguish Newtonian from non-Newtonian models without appealing to arbitrary conventions about metric representation. It is a proposal about frame individuation, not about the epistemic status of inertia.

Nevertheless, I argue that it supports a more ambitious reinterpretation: that inertia is a structural principle embedded in the symmetries of the theory. In this respect, it aligns with \cite{Earman1992-jn}'s SP principle framework, which emphasises the interdependence of dynamical and spacetime symmetries.

This reconceptualisation becomes particularly significant in the transition from Newtonian mechanics to GR, where Newton’s First Law is often said to be subsumed by the \textit{geodesic principle}. 
As I will argue, however, this move is deceptive: geodesic motion fails to represent the actual motion of bodies, and thus cannot recover the conceptual role played by the classical principle of inertia in this principled fashion.

\section{The Einstein's Law of Inertia: The Geodesic Principle} \label{geodesic}

This section turns to the relativistic setting of GR. The transition does not resolve the conceptual difficulties identified in the classical framework; instead, GR offers a generalisation, while further complicating the status of inertia.

In GR, gravity is not described as a force but is represented by the curvature of spacetime. Accordingly, departures from inertial motion are no longer attributed to gravitational forces, as in Newtonian mechanics, but to the geometry of the spacetime manifold. This geometric picture is not unique to GR; it is also present in Newton–Cartan theory (NCT), where gravity is likewise geometrised and free-falling bodies follow geodesics of an affine connection compatible with degenerate spatial and temporal metrics.\footnote{For detailed expositions of NCT, see \citet[\S3]{Earman1973}, \citet{Trautman1965-TRALOG,Trautman1967}, and \citet[\S4]{James_Read2023-mk}.}

The standard relativistic counterpart to Definition \eqref{INERTIA1} is as follows:

\begin{definition}
    \textbf{Inertial Motion (v.2)}: A body undergoes inertial motion if and only if it moves along a geodesic of the \textit{unique} Levi–Civita connection associated with the spacetime metric, \emph{i.e.} it is in\textit{ free fall} in a gravitational field. \label{INERTIA2}
\end{definition}

Definition \eqref{INERTIA2} makes explicit what is already implicit in standard relativistic practice: an object at rest on Earth's surface is \emph{not} in inertial motion, even if it would be so classified under the classical Definition \eqref{INERTIA1}.  For example, a rock resting on the ground remains at rest in the absence of external forces.
Yet such an object experiences a normal force opposing gravity. An accelerometer placed at rest on the ground registers a non-zero acceleration—precisely because the ground prevents the object from following its natural geodesic trajectory.
By contrast, a freely falling object, subject only to gravitation, exhibits no such acceleration and is deemed inertial in the relativistic sense.\footnote{This point is well established in the physics literature; see, for example, \citealp[p.67]{Wald93}.}

This formulation expresses the familiar claim that \textit{freely falling} bodies trace geodesics, which are often regarded as the \textit{natural trajectories} of motion—though I will challenge that designation in next sections.

Einstein explicitly endorsed this view when he introduced \lq{}the law of motion of General Relativity\rq{} as follows:
\begin{quote}
[\dots] a gravitating particle moves in a geodesic line. This constitutes a hypothetical translation of Galileo’s law of inertia to the case of the existence of \lq{}genuine\rq{} gravitational fields \cite[p.113]{Einstein1922}.
\end{quote}

Significantly, Einstein did not restrict this claim to infinitesimal bodies. He applied it to extended systems, including Mercury, whose perihelion precession he famously explained by modelling the planet as a point mass moving along a geodesic in curved spacetime.\footnote{See \cite{Einstein1916,Einstein1922}.} 
This use of the geodesic principle for a spatially extended, backreacting system raises deep conceptual and mathematical issues, which I will revisit in \S\ref{THEOREMS}.

Contemporary formulations typically elevate this idea to the status of a \textit{geodesic principle}: freely falling bodies move along geodesics of the spacetime metric. The geodesic principle is thus widely taken to express the relativistic analogue of inertia \citep{misner2017gravitation,Wald1984}.\footnote{Geodesic motion is also the standard characterisation of inertial motion in NCT. In that setting, the affine connection $\nabla^{NCT}$ is compatible with a temporal one-form field and a degenerate spatial metric. Freely falling particles follow geodesics of $\nabla_{NCT}$ defined by: $\nabla^{NCT}_{u^a}u^a=0$. For discussion on the geodesic principle in NCT see \cite{Weatherall2011}.\label{NCT}}

Importantly, Definition \eqref{INERTIA2} reflects a crucial conceptual shift. In GR, inertial motion—now identified with geodesic motion—is inherently \textit{local}, applying only along a given geodesic.
More precisely, inertial motion corresponds to the \textit{local flatness} of the Levi–Civita connection—that is, the ability to choose a reference frame such that the connection coefficients $\Gamma^a_{bc}$ vanish locally.\footnote{Strictly speaking, it is an abuse of notation to write $\Gamma^a_{bc}$ in abstract index notation, as these components only become meaningful once a coordinate basis is fixed. One should distinguish these coefficients from the abstract covariant connection operator $\nabla$.} However, due to spacetime curvature, these frames cannot in general be extended beyond infinitesimal neighbourhoods. Inertial motion thus characterises the local experience of a freely falling body.

Yet this interpretation demands care. Definition \eqref{INERTIA2} identifies inertial motion with geodesic motion relative to the Levi–Civita connection. But this connection can be non-zero even in the absence of curvature. That is, the geometry may be \textit{curvilinear} but not \emph{curved}.
In such cases, the Riemann tensor vanishes $R^{a}{}_{bcd}=0$, yet the connection coefficients $\Gamma^a_{bc}$ do not. The resulting inertial effects—so-called fictitious forces—arise not from spacetime curvature, but from the observer’s choice of a non-inertial frame.
Accordingly, the identification of inertial motion with geodesic motion does not presuppose curvature, but only the presence of a spacetime connection—curved or not—within which motion can be defined.

The distinction is crucial. In flat spacetime, inertial motion is globally definable, and deviations from it—when not due to real forces—reflect only the observer’s frame. 
These fictitious forces can be removed globally by transforming to an appropriate inertial frame. In this sense, inertial motion is not obstructed but merely misrepresented.

By contrast, in curved spacetime, curvature itself imposes limits on the extension of inertial frames: no global transformation can eliminate tidal effects encoded in the Riemann tensor. Inertial motion becomes an inherently local phenomenon, defined only in infinitesimal neighbourhoods where the connection can be flattened.

From Definition \eqref{INERTIA2}, one may now formulate a relativistic counterpart to \eqref{PIN1}:

\begin{definition}
    \textbf{PIN. v.4}: A body maintains inertial motion \textit{if and only if} its motion is determined by no interaction other than gravity. \label{PIN4}
\end{definition}

The phrase \lq{}no interaction other than gravity\rq{} must be understood carefully. It does not refer to the mere presence of spacetime curvature (i.e., a non-vanishing Riemann tensor),  but to the condition that the body’s dynamics is governed solely by the spacetime metric and its associated Levi–Civita connection.
In other words, the body’s Lagrangian contains no additional coupling terms—no electromagnetic fields, no internal propulsion, no interaction beyond gravity understood as what is encoded in the connection.

Under this interpretation, the biconditional in Definition \ref{PIN4} is justified. If no non-gravitational interaction determines a body’s motion, then the general relativistic equations of motion entail that it follows a geodesic. Conversely, if a body follows a geodesic, this implies that no external interaction is acting on it: it evolves freely under the influence of the connection alone.

This accommodates both curved and flat spacetimes. In Minkowski space, a body might move geodesically not because gravity is the \textit{only} interaction, but because \textit{there is no interaction at all}.
Importantly, the biconditional does \textit{not} imply that geodesic motion requires spacetime curvature: the geodesic equation is well-defined even in flat spacetimes, where $R^a_{bcd}= 0$. 
What matters is not the curvature tensor, but the affine structure provided by the connection. Geodesic motion reflects the dynamical absence of non-gravitational couplings, not the presence of gravitational curvature.

Although this formulation removes the circularity of earlier classical definitions, it veers toward triviality. Since geodesics are \textit{defined} as the curves traced by free-falling bodies, the assertion that bodies move inertially when acted upon only by gravity simply reiterates the definitional content of GR’s geometry.  It does not explain \textit{why} free bodies move as they do. 
More significantly, as I will argue in §\ref{THEOREMS}, this apparent definitional clarity masks a deeper conceptual tension. While geodesic motion is geometrically well-defined, it lacks any clear referent. There is, properly speaking, \textit{no such thing} as the geodesic motion of a body.

Following the rationale adopted in the classical setting, one may be introduce the notion of an \textit{inertial reference frame}. Importantly, in GR we can only introduce a \textit{local} inertial reference frame.
Analogously to Definition  \eqref{INRF}, we may state:

\begin{definition}
    \textbf{(Local) INRF (v.2)}: A \textit{local} inertial reference frame provides a \textit{local} standard of space and time measurement, defined by parameters $\{x^{I}\}$, in which the geodesic equation \begin{equation}
\frac{D^2 x^a}{d\tau^2} = \frac{d^2 x^a}{d\tau^2} + \Gamma^a_{bc} \frac{dx^b}{d\tau} \frac{dx^c}{d\tau} = 0
\end{equation} reduces to \begin{equation}
\frac{D^2 x^I}{d\tau^2} = \frac{d^2 x^I}{d\tau^2}= 0 \quad (\text{with }\Gamma^I_{JK}=0\text{ locally}).
\end{equation} In such a frame, any deviation from this equation arises solely from non-gravitational causes. \label{INRF2}
\end{definition}

This definition underscores the above mentioned central insight of GR: inertiality is fundamentally \textit{local}. It holds only where the connection can be rendered locally flat.
As I will argue in §\ref{EQUIVALENCE}, no physically realisable inertial frames exist over \textit{infinitesimal} regions of spacetime. This undermines the viability of local inertial frames in the operationalist sense—whether understood weakly, as empirically anchored, or strongly, in the sense of \cite{bridgman1936nature}, where operational meaning depends on explicit measurement procedures.\footnote{An alternative, \textit{non-operational} definition of a \textit{global} inertial reference frame in GR is given in \cite{Earman1973}, where it is defined by a timelike vector field $X$ satisfying: (i) $X$ is a Killing field, (ii) its integral curves are hypersurface orthogonal, and (iii)  the proper time between hypersurfaces along $X$ is constant.
As a consequence, in this frame the metric to take the ultrastatic form $ds^2=-dt^2+g_{ij}dx^idx^j$, with \textit{lapse function} (relating coordinate time to proper time) equal to unity \citep{Wald1984}.
However, such global inertial frames exist only in ultrastatic spacetimes, where geodesic observers are globally synchronisable and both gravitational time dilation and frame-dragging are absent. This excludes, for instance, all rotating observers \citep{Landau1987-fh}. The rarity of such spacetimes reinforces the point that inertial frames in GR are, in general, \emph{only locally} definable.}

It is also essential to note that a local \textbf{INRF} is not necessarily a \textit{synchronous} frame \citep{Landau1987-fh, bamontithebault}. That is, the proper time $\tau$ measured by a freely falling observer need not coincide with the clock parameter $x^4$ of the frame adapted to that observer.
This reflects the general fact that geodesics may be parametrised by an \textit{arbitrary} affine parameter $\lambda = \alpha\tau + \beta,$ $(\alpha \neq 0,\, \beta \in \mathbb{R})$, which has no direct interpretation as proper time measured by some clock (see the \textit{clock hypothesis} in \citealp{malament2012topics}).

Crucially, Definition \eqref{INRF2} does not entail that \textit{every} reference frame in a purely gravitational setting is inertial. To illustrate this, consider three observers situated in the exterior region of a Schwarzschild black hole of mass $M$:

\begin{itemize}

\item \textbf{Alice}  is in radial free fall. Along her worldline, the geodesic equation reduce to $\frac{d^2 x^I}{d\tau^2} = 0$, and the connection coefficients vanish: \( \Gamma^I_{JK} = 0 \) . Alice's frame $\{x^I\}$ is a \textit{local inertial frame} adapted to her geodesic.

\item \textbf{Bob} hovers at a fixed Schwarzschild radius \( r_0 \) (as measured by an observer at infinity, named Carl), sustained by a rocket. His proper 4-acceleration, whose spatial magnitude is measurable via an onboard accelerometer, counteracts the black hole’s pull. In this respect, Bob resembles a body resting on the Earth’s surface. 
His frame  \( \{y^I\} \) is non-inertial, and this is revealed in the geodesic equation he assigns \textit{to Alice}'s motion:
\begin{equation}
    \frac{d^2 y^I}{d\tau^2} + \Gamma^I_{JK} \frac{dx^J}{d\tau} \frac{dx^K}{d\tau} = 0.
\end{equation}
Here, the non-vanishing connection coefficients  \( \Gamma^I_{JK} \neq 0 \)  encode fictitious forces in Bob’s frame—reflections of his own proper acceleration. The non-inertial character of his frame stems not from spacetime curvature \textit{per se}, but from the internal interaction (the rocket) that maintains his stationary position.

\item \textbf{Carl} is situated at spatial infinity, where spacetime is asymptotically flat. He defines a \emph{global inertial frame} \( \{z^I\} := \{t_C, r_C, \theta_C, \phi_C\} \), corresponding to standard Schwarzschild frames.\footnote{In most presentations, Schwarzschild coordinates are not conceived as reference frames. Since Carl is an asymptotic observer, his physical role is idealised. So, $\{z^I\}$ should be interpreted as a (global) \textit{idealised reference frame} in the sense of \cite{Bamonti2023}. As Bamonti observes (\S4), such idealised frames are often conflated with coordinate systems because they serve functionally similar roles.} 
In Carl’s frame, Alice’s radial free-fall satisfies:
\begin{equation}
    \frac{dr_C}{dt_C} = -\left(1 - \frac{2M}{r_C}\right)\sqrt{\frac{2M}{r_C}},
\end{equation}
while Bob’s trajectory is simply \( dr_C/dt_C = 0 \), since his position is fixed at \( r_C = r_0 \).
\end{itemize}

This example underscores a key conceptual point: non-inertial effects persist even in pure gravity scenarios, provided the observer’s own dynamics is sustained also by interactions that are not gravitational in nature. Bob’s situation is a case in point.
Although he is subject only to the gravitational field of the black hole, his motion is not governed solely by that field. The rocket’s thrust constitutes a non-gravitational interaction which introduces additional terms in the Lagrangian beyond the purely gravitational one.

In detail, Bob's dynamics cannot be derived from the minimal coupling of a free particle to the spacetime metric. Let's consider Carl's frame for the sake of simplicity. While Alice’s trajectory extremises the action
\begin{equation}
    S_A=\int{\sqrt{-g_{IJ}\dot{z}^I \dot{z}^J}d\lambda},
\end{equation}
Bob's trajectory must be derived from the action

\begin{equation}
    S_B=\int{\Big[\sqrt{-g_{IJ}\dot{z}^I \dot{z}^J}+A_I(\lambda)\dot{z}^I\Big]d\lambda},
\end{equation}
where $A_I(\lambda)$ represents a non-gravitational force term from controlled propulsion. This term  has no geometric origin and encodes the thrust direction and magnitude along Bob's path. The resulting Bob's Euler–Lagrange equations take the form
\begin{equation}
    \cfrac{Du^I}{d\tau}=a^I(\tau),
\end{equation}
where $u^I$ is Bob's 4-velocity and $a^I$ is Bob's proper 4-acceleration.

This contrasts with Alice’s case, for whom $a^I=0$ and $Du^I/d\tau=0$. Accordingly, even though Bob’s trajectory can be described within GR using a non-inertial coordinate system and non-vanishing Christoffel symbols, his actual motion cannot be attributed solely to the gravitational interaction encoded in the Levi–Civita connection.

Consequently, Bob's motion does not falsify PIN (v.4) precisely because his dynamics include such a non-gravitational interaction. 
The fact that the rocket’s thrust can be geometrised, i.e. absorbed into into non-zero Christoffel symbols, does not render it gravitational in the relevant sense. 
The key diagnostic is the Riemann tensor: while Christoffel symbols can appear in both inertial and non-inertial frames, true gravitational interaction is inseparable from spacetime curvature. 
Bob’s trajectory, maintained by a rocket in a centrally curved geometry, mimics the effects of a \textit{homogeneous} field, but it is not gravitationally determined (see also below \S\ref{EQUIVALENCE}).

Accordingly, Definition \ref{PIN4} remains valid: a body moves inertially if and only if its motion is determined solely by no interaction other than gravity, that is if and only if there are no additional couplings in the Lagrangian.

In analogy with Definition \eqref{PIN3}, Definition \ref{PIN4} may be restated in terms of \textbf{INRFs} (but here, local):
\begin{definition}
    \textbf{PIN. v.5}: \textit{Relative to} a \textit{local} inertial reference frame (as defined in \ref{INRF2}), a a body maintains its geodesic motion: $d^2x^I/d\tau^2=0$.\label{PIN5}
\end{definition}

Despite its formal clarity, Definition \ref{PIN5} brings back the problem we encountered in the classic case: circularity.
It defines inertial motion by reference to local inertial frames—yet those frames, in Definition \ref{INRF2}, are themselves defined as the ones in which bodies move inertially. This mutual dependence empties the principle of its explanatory force: it tells us that a body moves inertially when its motion matches the behaviour that defines the very frame being used.

A further conceptual vulnerability of Definition \ref{PIN5} lies in its reliance on \textit{locality}.
The very possibility of defining inertial frames depends on the capacity to render the connection coefficients locally zero, typically justified by appeal to the Equivalence Principle, which asserts that gravitational effects can be cancelled in sufficiently small, \textit{local} neighbourhoods.
But what exactly does \textit{local} mean in this context and what does this reliance imply for the physical significance of geodesic motion?

These concerns are not merely semantic. As I will argue, if inertial motion is definable only within vanishingly small regions, then it ceases to be a physically realisable mode of motion. It becomes instead a formal artefact of differential geometry. As such, the physical content of the geodesic principle grows increasingly obscure—particularly when applied to bodies with finite extension, internal structure, or non-negligible backreaction.
The next subsection turns to these questions.

\subsection{A First Challenge From the Equivalence Principle: What is Local?}\label{EQUIVALENCE}

According to Definition \eqref{INERTIA2}, inertial motion is identified with the geodesic motion of a curved Levi-Civita connection $\nabla_g$, compatible with a curved spacetime metric $g_{ab}$.
This connection is said to be \textit{locally flat}, in the sense that its components can always be made to vanish at a point by an appropriate choice of frame.
This reflects the standard claim that \lq{}spacetime is locally flat\rq{} and that \lq{}special relativity holds locally\rq{}.\footnote{See \citealp{Brown2016,Read2018,GomesSR,Fletcher2023-FLETLV,Teh2024} for discussions on the validity and scope of these claims. \label{fnlocalsr}}
In practice, this local inertial structure is often taken to mean that \textit{gravitational effects vanish locally}, as typically justified by appeal to the Equivalence Principle (EP) \citep[§1.4]{ferrari2020general}.\footnote{This paper refers specifically to the \textit{Strong Equivalent Principle}. This must be distinguished from th  \textit{Weak Equivalence Principle} (WEP), which concerns the equivalence of inertial and gravitational mass, and from the \textit{Einstein Equivalence Principle} (EEP), which addresses the identity of gravity and inertia as an ontological claim. The SEPasserts the local validity of special relativity—i.e., the existence of local frames in which freely falling bodies behave as if no gravity were present. See  \cite{Lehmkuhl2022-LEHTEP} for a thorough analysis.}

Earman and Friedman once remarked, with some irony, that \lq{}\lq{}there are almost as many interpretations of this principle as there are authors of books on relativity theory\rq{}\rq{} \citep[p. 329]{Earman1973}. 
Among the most influential these is the so-called \textit{infinitesimal} or \textit{local} equivalence principle, often attributed to \cite{pauli2013theory} and discussed in detail by \cite{Norton1985}.
It asserts that gravitational motion is \textit{locally} indistinguishable from inertial motion, and the gravitational effects can always be cancelled at a point.

Steven Weinberg offers a standard expression of this idea:
\begin{quote}
Locally, the effects of a gravitational field are equivalent to those experienced in a non-inertial reference frame. Thus, gravity can be cancelled locally and this possibility defines local inertial frames where $\Gamma^a_{bc}=0$. \citep[p.68]{weinberg1972gravitation}\
    
\end{quote}
The Equivalence Principle (EP) is not an incidental feature of geodesic motion—it is constitutive. 
As shown in standard derivations (cf. \citealp[\S1.6]{ferrari2020general}), the geodesic equation arises precisely from the demand that, in a locally inertial frame—guaranteed by the EP—the equations of motion reduce to their special relativistic form: bodies move in straight lines at constant speed. When transforming to a general frame, derivatives of the metric introduce non-null connection terms, leading to the familiar geodesic equation. 
In this sense, the geodesic principle is a formal consequence of the Equivalence Principle.

It is important to recall, however, that the EP was not originally cast in such local terms. In its earliest formulation—Einstein’s famous \lq{}happiest thought\rq{}—the emphasis was \textit{global} rather than infinitesimal. 
Einstein considered the equivalence between a \textit{homogeneous} gravitational field and the \textit{uniform} acceleration in flat spacetime.
This insight later gave rise to what might be termed the \textit{vulgata} version of the EP: that gravity and fictitious, non-inertial forces are \lq{}two sides of the same coin,\rq{} and that the gravitational field has only a relative existence, since in some reference frame it can be transformed away, restoring the inertiality condition of motion $ \frac{d^2x^a}{d\tau^2}=0 $.

Historically, this transition from a global to a local EP was cemented by Pauli, who generalised Einstein’s reasoning to \textit{arbitrary} gravitational fields and \textit{arbitrary} accelerations.
Pauli argued that \textit{any} gravitational field could be nullified \textit{at a point} by a suitable diffeomorphism—just as \textit{any} fictitious force could.
This yields the modern infinitesimal form of the EP: gravity can always be \lq{}transformed away\rq{} in an \textit{infinitesimal} region of spacetime, just as the components of the Levi-Civita connection can be made to vanish locally, rendering gravity \textit{locally} equivalent to a non-inertial force.\footnote{As a historical aside, a rough precursor to Einstein’s original global EP appears in Newton's \cite{newton1687principia}'s \textit{Principia}, Corollary VI. Here, Newton introduces accelerated systems that behave, for practical purposes, \textit{as if at rest or in uniform motion}. These \lq{}quasi-inertial\rq{} systems prefigure the idea that certain forms of acceleration may be physically indistinguishable from gravitational effects.
\begin{quote}
    \textbf{Corollary VI to the laws of motion}:
If bodies are moved in any way among themselves, and are urged
by equal accelerative forces along parallel lines, they will all
continue to move among themselves in the same way as if they
were not acted on by those forces (ibid., p. 20.)
\end{quote}}

Beneath the rhetorical power of the EP lies a conceptual ambiguity. The very notion of \textit{locality}—central to its formulation—admits at least two distinct interpretations:
\begin{itemize}
\item \textbf{Local at a point:} This refers to an infinitesimal neighbourhood around a single manifold point on the manifold—conceptually treated as shrinking to the point itself, and which has zero Lebesgue measure in the ambient spacetime. In coordinate-based terms, this notion underlies the use of \textit{Riemann normal coordinate}s (adapted to a point).
\item \textbf{Local along a geodesic:} This refers to an infinitesimal \lq{}tubular\rq{} neighbourhood surrounding a timelike geodesic. While also of zero Lebesgue measure, the geodesic itself admits an affine parametrisation, resulting in a finite or even unbounded proper length. In coordinate-based terms, this notion underlies the use of \textit{Fermi normal coordinates} (adapted to a geodesic) \citep{fermi1922sopra}. 
\end{itemize}

Both notions of locality are mathematically well-defined, but neither accommodates physically possible systems. 
Thus, the EP, on either reading, is demoted to a \textit{useful construct} masquerading an operational one.  It expresses a mathematical property of the connection, but cannot be instantiated by any admissible matter configuration in GR.\footnote{This is the infinitesimal counterpart of \cite{Hilbert_1984}'s remark that \lq{}\lq{}the infinite is nowhere to be found in reality, no matter what experiences, observations, and knowledge are appealed to.\rq{}\rq{}. Just as the infinitely large is a mathematical construct, so too is the infinitely small.} 

The geodesic principle inherits this same limitation. The problem is not merely technical—it is ontological.
Any adequate account of inertial motion in GR must eventually confront the status of geodesics themselves.
But geodesics are defined only in infinitesimal domains, and satisfied only by bodies with no extension, no structure, and no backreaction.

As the next section (\S\ref{THEOREMS}) will make clear, geodesics are not even legitimate idealisations. They are, at best, \textit{formal} markers of how motion would proceed given the structure of the connection—markers that fail to correspond to any real or fictitious system compatible with Einstein’s equations.

\subsection{Summary: Circularity, Triviality, and the Fate of Inertial Motion}

The definitional landscape of inertial motion—across both classical and relativistic frameworks—reveals a persistent tension. In the classical case, most formulations of the Principle of Inertia tend to collapse into \textit{circularity}: they presuppose inertial motion in the very concepts they invoke to define it, whether force, inertial reference frames, or isolation.

In the relativistic setting, the problem shifts: some formulations (such as PIN (v.4)) avoid circularity but collapse into \textit{triviality}—reducing the Principle of Inertia to a restatement of geodesic structure. Others, such as PIN (v.5), are no better than their classical counterparts: they reintroduce \textit{circularity} by defining geodesic motion in terms of a local inertial reference frame—one which is itself characterised by geodesic motion. Once again, the explanatory ambition of the principle collapses. What was meant to be a physical criterion for distinguishing types of motion dissolves into either circularity or definitional vacuity.

This diagnosis is reinforced by the analysis of the Equivalence Principle in §\ref{EQUIVALENCE}.
The idea that gravity can be \lq{}cancelled\rq{} in a local inertial frame—and that geodesic motion thereby counts as locally inertial—rests on a mathematically valid but physically fragile notion of \textit{locality}.
Although the Levi-Civita connection can always be made to vanish at a point or along a geodesic, this does not entail that gravitational effects, encoded in the Riemann tensor, are absent, \textit{even locally}.
The inference from geometric locality to physical inertiality breaks down.

This failure is not merely technical. It indicates a deeper disconnect between the formalism of GR and the empirical content that a principle of inertia is meant to provide. 
Geodesic motion, while often taken to define inertial motion in GR, cannot model the motion of actual bodies. 
As I will show in the next section, it cannot even be recovered as a valid limit of  matter systems obeying the EFEs. So, strictly speaking, inertial motion is not only empirically insignificant, it is also not derivable within GR.
The principle that once codified the structure of natural, free motion bodies of in GR now seems to be nothing more than a geometric detail.

This sets the stage for the central interpretive move of the paper, thoroughly developed in the next section: geodesic motion is neither an approximation to the behaviour of real bodies, nor a property of an idealised surrogate, since no system in GR, real or fictitious, can possess its defining properties.
The result is a subtle but decisive shift in interpretational status: inertial motion, understood as geodesic motion, is not the motion of \textit{any} real or even idealised body. 
It is a property of the Levi–Civita connection itself, not of anything that could instantiate it. 
Inertial motion, so defined, becomes a \textit{useful construct}——a geometrical artefact unmoored from the space of dynamically possible systems (whether real, possibly real, or fictitious).

In its place, I propose a different organising concept: the \textit{natural motion} of bodies. Unlike inertial motion, natural motion is grounded in dynamically admissible regimes. It admits a layered structure of approximations, \textit{each valid for a class of systems} with particular physical features—internal structure, spatial extension, or gravitational feedback. 
In each regime, the equations of motion arise from consistent applications of GR’s dynamical laws. 
As such, they form a hierarchy of genuine approximation schemes. Geodesic motion belongs to none of them.

Since antiquity, natural and inertial motion have often been treated as coextensive: to move naturally is to move inertially, and vice versa. What follows challenges this equivalence. The two notions should be carefully distinguished. Natural motion is dynamically grounded and admits physical referents; inertial motion, as geodesic motion, does not.

In the next section, I turn to the various attempts to \textit{derive} the geodesic principle from within general relativity itself. These derivations—ranging from the Geroch–Jang and Ehlers–Geroch theorems to the Einstein–Grommer and Geroch–Traschen approaches—clarify why geodesic motion fails to serve even as a consistent idealisation. Each case illustrates a different kind of failure. Together, they support the reclassification of inertial motion as a formal artefact, lacking both explanatory and representational power. This will mark the transition to the second half of the paper (\S\ref{EXTENSION}-\ref{naturalmotion}), where I develop the layered account of natural motion as a dynamically grounded alternative.

\vspace{1em}

\noindent
The following Table \ref{TAB1} summarises the status of each definition introduced thus far:

\begin{table}[h!]
\centering
\renewcommand{\arraystretch}{1.3}
\begin{tabular}{|l|>{\raggedright\arraybackslash}p{5cm}|>{\raggedright\arraybackslash}p{3.1cm}|>{\raggedright\arraybackslash}p{5.1cm}|}
\hline
\textbf{Definition} & \textbf{Description} & \textbf{Circular?} & \textbf{Trivial?} \\
\hline
\textbf{Def.~\ref{INERTIA1}} & Inertial motion = uniform motion or rest & No & No (but vague): uniform motion presupposes a privileged class of frames\\
\textbf{Def.~\ref{PIN1}} & PIN (v.1): no external force acts & Yes & No \\
\textbf{Def.~\ref{PIN2}} & PIN (v.2): sufficient distance from other bodies & Yes & No \\
\textbf{Def.~\ref{INRF}} & INRF (v.1): defined via Newton’s laws & Yes & No \\
\textbf{Def.~\ref{PIN3}} & PIN (v.3): Newton's laws hold in INRF (v.1) & Yes (inherited from Def.~\ref{INRF}) & No \\
\textbf{Def.~\ref{INERTIA2}} & Inertial motion (v.2): geodesic motion in GR & No & Yes: no explanation of why geodesics count as inertial \\
\textbf{Def.~\ref{PIN4}} & PIN (v.4): a body is inertial if it follows a geodesic & No & Yes (inherited from Def.~\ref{INERTIA2}) \\
\textbf{Def.~\ref{INRF2}} & Local INRF (v.2): where geodesic motion appears as uniform & Yes. The frame is defined by appeal to geodesic motion (Def. \ref{INERTIA2}). & Yes (similar to Def.~\ref{INERTIA1}): no explanation of why the frame is inertial \\
\textbf{Def.~\ref{PIN5}} & PIN (v.5): inertial motion is geodesic motion in a local INRF (v.2) & Yes (inherited from Def.~\ref{INRF2})  & Yes (inherited from Def.~\ref{INRF2}) \\
\hline
\end{tabular}
\caption{Definitions of inertial motion and the Principle of Inertia and their shortcomings.}
\label{TAB1}
\end{table}

\section{The Limits of the Geodesic Principle}\label{THEOREMS}

\begin{quote}
\textit{If you wish to learn from the theoretical physicist anything about the methods which he uses, I would give you the following piece of advice: Don't listen to his words, examine his achievements. For to the discoverer in that field, the constructions of his imagination appear so necessary and so natural that he is apt to treat them not as the creations of his thoughts but as given realities.}

\rightline{\citealp{Einstein_1934}}
\end{quote}
The preceding analysis has already cast doubt on the conceptual integrity of the geodesic principle. Although its formal definition—motion along Levi-Civita geodesics of a Lorentzian metric— is clear, its significance as an empirical principle appears to be devoid of substance.
Definitions that equate inertial motion with geodesic motion were shown to suffer from circularity or triviality. Moreover, the very concept of geodesic motion relies on a notion of locality that collapses when extended to physically realistic, spatially extended systems.

This section advances the critique. I argue that geodesic motion cannot serve as a description of how real, free bodies move, nor can it be justified as a limiting case of their behaviour.
 It neither approximates the dynamics of any admissible system in GR, nor represents the behaviour of any idealised surrogate compatible with EFEs.
The kinds of bodies to which the geodesic principle might apply—infinitesimal, structureless, non-backreacting—are not merely unrealistic: in many cases, they are dynamically inconsistent with the theory itself. Where they can be formally constructed, they either violate the field equations or fail to exhibit the defining property of geodesic motion. 
\textit{No real target system can be approximated by them; no consistent idealisation can instantiate them}. 
Geodesic motion, although geometrically well-defined, has no referent whatsoever. 
What is at stake is not merely its empirical applicability, but its theoretical legitimacy.

This diagnosis is anticipated in \cite{Tamir2012}’s  three-pronged critique of the geodesic principle:
\begin{quote}
Specifically, I argue for the following three claims.
First, [\dots] massive bodies are [n]ever guaranteed to follow geodesic paths.
Second, [\dots] extended massive bodies generically deviate from uniformly geodesic paths.
[\dots] Third, thanks to certain mathematical theorems concerning distribution theory, alternative representations of massive bodies as unextended "point" particles must result either in precluding the possibility of coupling the particle to the spacetime metric in a way that is coherent with Einstein’s field equations or in having to excise the particle (and its would-be path) from spacetime entirely.
This three-pronged argument reveals that [\dots] the geodesic principle in such a way requires that either the gravitating body is not massive, its existence violates Einstein’s field equations, or it does not exist within the spacetime manifold at all (let alone along a geodesic) (ibid., p.137-138).
\end{quote}

Tamir’s analysis is physical: the geodesic principle fails to describe the motion of any body whose dynamics are governed by GR. 
Either the body must violate the EFEs, vanish entirely, or lie outside the manifold itself.

In this section, I refine and extend that critique using a framework introduced by \cite{Norton2012}.  Norton distinguishes two criteria by which a physical property—such as geodesic motion—might be recovered from a limiting process.

An \textit{Idealisation} succeeds only if the limiting behaviour, the \textit{limit property}, is carried by a coherent\textit{ limit system}. 
An \textit{Approximation}, by contrast, succeeds only if the property in question approximates the behaviour of an existing real target system.

\textit{Geodesic motion satisfies neither criterion}. 

Norton identifies two main modes of failure of an idealisation: either (i) \textit{the limit system does not exist}, or (ii) \textit{the limit system exists but does not bear the limit property (geodesic motion)}.
The argument that follows is structured explicitly around these two modes of failure of a body following geodesic motion.

Norton also recognise that also an approximation may becomes ‘pathological’. For example, when the mathematical procedure that generates it produces a description that is inconsistent, indeterminate, or meaningless for the system being approximated.
In particular, he identifies two main ways approximations can fail:
\begin{enumerate}
    \item \textbf{Tolerance Violation:} This occurs when the approximation error exceeds the acceptable level of inaccuracy for the specific context, which can vary greatly depending on the situation. For example, a 10\% error might be acceptable in one context, but not in another. This renders the approximation too inaccurate to be useful.
    \item \textbf{Pathological Tracking:} This occurs when the method used to generate the approximation  fails to produce a coherent, stable, or adequately representative description of the target system. In such case, the inaccurate propositional description ceases to reliably \lq{}track\rq{} the behaviour of the target system.

\end{enumerate}

I suggest focusing on a specific example of case (2), which I will use below to demonstrate that geodesic motion is not a reliable approximation in GR:
\begin{itemize}
    \item[2.1] \textbf{Off-Shell Failure:} This occurs when the mathematical procedure constituting the approximation \textit{violates} the theory’s own dynamical equations—ceasing to be even an approximate solution within the theory, and thus structurally invalid. Importantly, this should \textit{not} be confused with the case in which a solution cannot bear the well-constructed approximated property. The failure concerns the procedure itself.
\end{itemize}

Following Tamir, I divide derivations of the geodesic principle into two broad classes:

\begin{itemize}
    \item \textbf{Limit proofs}, which consider sequences of stress–energy tensors confined to ever-smaller spatial regions surrounding a timelike curve. These proofs aim to show that if such a sequence converges appropriately, the limiting curve must be a geodesic. The conclusion is that arbitrarily small bodies follow geodesics.
    \item \textbf{Singularity proofs}, which invoke singularities—typically bodies that lie outside the spacetime manifold, or are replaced with curvature divergences. These do not involve limiting procedures and rely instead on pathological constructions.
\end{itemize}

While Tamir uses this classification to argue that no derivation justifies geodesic motion for real, massive bodies,  my aim is to show something stronger: that geodesic motion is neither an approximation nor an idealisation. It is a formal artefact of the connection, not a representational or dynamical feature of any body, real or ideal.

To make this case, I examine four canonical derivations of the geodesic principle:
\begin{enumerate}

\item The \textit{Geroch–Jang theorem}, which is the most influential attempt to derive the geodesic principle from within the theory. 
It purports to show that bodies with smooth, conserved, compactly supported stress–energy must move along geodesics. But the result applies only in a fixed background geometry: the matter distribution is allowed to have non-zero stress–energy while being prevented from perturbing the geometry. This amounts to assuming the test-body limit \textit{without justification}.  That is, the assumption of zero backreaction is introduced ad hoc and not justified by any dynamical argument from the EFEs. The geodesic principle here is not derived, but presupposed.  As such, the theorem functions not as a limit proof, but as a constraint on admissibility and, as such, does not suggest anything about whether geodesic motion is an idealised motion or not.  However, it does suggest that geodesic motion cannot be used to approximate the motion of a real system in GR.

\item The \textit{Ehlers–Geroch theorem}, a true limit proof, attempts to recover geodesic motion from a sequence of matter-filled spacetimes whose metrics converge to a background geometry. But in the limit, either the stress–energy vanishes or the field equations are violated. 
The limit system exists, but it does not bear the geodesic (limit) property—an instance of Norton’s second failure mode: \textit{limit property and limit system disagree}. Also, geodesic motion fails as an approximation because at no stage of the limit proof does geodesic motion serve as an even approximately valid description of target bodies' dynamics.

\item The\textit{ Einstein-Grommer strategy}, a singularity proof, seeks to derive geodesic motion by excising the body from the manifold. This is not a limiting procedure; the body and its worldline are simply removed from the beginning. Hence, no target system can be approximated by this strategy—excluding approximation altogether.

\item Finally, the \textit{Geroch–Traschen theorem}, also a singularity result, proves that no distributional stress–energy source supported on a curve can satisfy the EFEs. This blocks from the outset the construction of a consistent limit system altogether, instantiating Norton’s first failure mode: \textit{there is no limit system}.

\end{enumerate}

The conclusion is stark: geodesic motion is not the trajectory of \textit{any} body—real or idealised—but rather a \textit{useful yer uninstantiable construct}.  It is a formal property of the Levi–Civita connection, not a dynamically realisable trajectory.

\subsection{The Geroch-Jang Theorem.}\label{gerochjang}

The Geroch–Jang theorem is often presented as a major justification for the geodesic principle within GR \citep{Geroch1975}. 
Its conclusion is commonly interpreted to support the idea that free-falling bodies of positive mass must follow geodesics, and that this behaviour is not postulated but rather derived from the structure of the theory itself. 
Some authors have extended this line of reasoning beyond GR, treating the result as a theorem about the motion of matter in any relativistic theory with appropriate geometric structures.
\footnote{\cite{Weatherall2019}  argues that the EFEs play no essential role in the \textit{geodesic theorem}. The principle qualifies as a theorem insofar as it follows, under appropriate assumptions, from the geometric structure and its role in governing matter dynamics—regardless of the specific gravitational field equations.  This perspective allows for analogous derivations in Newton–Cartan theory \citep{Weatherall2011} and other relativistic frameworks. For a contrasting view, which insists on the special role of GR in \textit{explaining} geodesic motion dynamically, see \cite{Brown2005-kq}; see also \cite{Sus2014} and \cite{Samaroo2018}. \label{fngeodetictheorem}}

But this interpretive tradition overstates what the theorem delivers. As Tamir has convincingly argued, the Geroch–Jang result does not establish geodesic motion as a consequence of Einsteinian dynamics, nor does it show that real or test bodies follow geodesics.
What it proves is more limited: that geodesic motion may be attributed to a curve, provided certain restrictive conditions are satisfied—conditions whose physical significance is unclear and whose dynamical status remains ungrounded.  In short, the theorem \textit{presupposes}, rather than derives, the test-body regime for which geodesic motion is valid.

Strictly speaking, test bodies are systems with vanishing stress–energy: they produce no gravitational field and evolve in a fixed background geometry satisfying the vacuum Einstein equations. The Geroch–Jang construction does not meet this condition.

To clarify the structure of the result, it is useful to state the theorem in formal terms (see \citealp[p.22, Theorem 3.1]{Weatherall2016-WEAIME}; see also \citealp[p.146]{malament2012topics}):

\begin{thm}\label{GJ}
Let $(\mathcal{M},g_{ab})$ be a relativistic spacetime, and suppose $\mathcal{M}$ is oriented. Let $\gamma:I\rightarrow \mathcal{M}$ be a smooth imbedded curve.  Suppose that given any open subset $O$ of $\mathcal{M}$ containing $\gamma[I]$, there exists a smooth symmetric field $T^{ab}$ with the following properties.
\begin{enumerate}
\item \label{sdec} $T^{ab}$ satisfies the \emph{strengthened dominant energy condition}, i.e., given any timelike covector $\xi_a$ at any point in $\mathcal{M}$, $T^{ab}\xi_a\xi_b\geq 0$ and either $T^{ab}=\mathbf{0}$ or $T^{ab}\xi_a$ is timelike;
\item \label{cons}$T^{ab}$ satisfies the \emph{conservation condition}, i.e., $\nabla_a T^{ab}=\mathbf{0}$;
\item \label{inside}$\text{supp}({T^{ab})}\subset O$; and
\item \label{non-vanishing}there is at least one point in $O$ at which $T^{ab}\neq \mathbf{0}$.
\end{enumerate}
Then $\gamma$ is a timelike curve that can be reparametrized as a geodesic.
\end{thm}

This result has often been interpreted to mean that \textit{arbitrarily small bodies of positive mass must follow geodesics}. 
 But this interpretation is misleading. It rests on a reading that clashes with the dynamical structure of GR. Three conceptual tensions emerge from the assumptions and consequences of the theorem.

\begin{description}

\item[(T1) \textbf{Stress-energy without backreaction.}] 

The key assumption of the theorem is that for \textit{any} open neighbourhood $O$ around the curve $\gamma$, however small, one can construct a smooth symmetric stress-energy field $T^{ab}$ supported entirely within $O$. 
By choosing a nested sequence of such neighbourhoods $(O_i)_{i\in\mathbb{N}}$ that shrink around $\gamma$ as $i\to\infty$, one obtains a sequence of stress-energy tensors $\underset{\tiny i}{T^{ab}}$ whose spatial support becomes arbitrarily small and converges to the curve.. These are the so-called \textit{Geroch-Jang particles} (GJ-particles): smooth compactly supported distributions of non-zero energy–momentum, increasingly localised near $\gamma$.\footnote{At first glance, the use of a sequence  $\underset{\tiny i}{T^{ab}}$ suggests a limiting procedure: the curve $\gamma$ acquires geodesic status as the limiting trajectory of shrinking matter configurations. But this impression is mistaken.
While a limit property (geodesicity of $\gamma$) is defined, there is no attempt to construct a  \emph{limit system} that includes the shrinking matter as a dynamically consistent source.
The failure here is not of Norton’s first or second kind, but of a different sort: the construction does not even engage the dynamical content of the theory.}

The core tension is this: if each GJ-particle represents real matter—however localised—then it ought to source a metric perturbation via the Einstein equations. Since the support of each  $\underset{\tiny i}{T^{ab}}$ shrinks \textit{but never vanishes}, the associated perturbation never vanishes, no matter how small the support becomes. The theorem, however, proceeds by ignoring this effect: the background geometry remains fixed throughout. 
This is not a well-constructed test-body limit—it is a formal artefact. The stress–energy tensor is present but dynamically inert.\footnote{Moreover, as \citet[p.22]{Weatherall2016-WEAIME} notes:
\begin{quote}
a principal difficulty in trying to derive the geodesic principle as a theorem [that is, as a consequence of dynamics and not as an independent postulate] concerns a kind of ontological mismatch between the geodesic principle and the rest of general relativity: namely, general relativity is a field theory, whereas the geodesic principle is a statement concerning point particles.\rq{}\rq{} (my aside).
\end{quote}
}
In this sense, the theorem proves only that geodesic motion can be \emph{assigned} to a curve in a fixed background—\textit{not} that it emerges dynamically from the theory’s equations.

These limitations motivated the refinement offered by the Ehlers--Geroch theorem, which attempts to recover geodesic motion while accounting for the backreaction of the matter fields. That result will be the focus of the next subsection.

\item[(T2) \textbf{Conservation and the Bianchi identities.}] 

One might attempt to defend the theorem’s conclusion by appealing to the conservation condition \(\nabla_a T^{ab} = 0\), which appears as a premise in the theorem. This condition is sometimes taken to imply that matter must follow geodesic paths. However, this inference is flawed.
In GR, $\nabla_a T^{ab} = 0$ is indeed guaranteed \emph{if}  $T^{ab}$ arises as the source of a metric via the EFEs, by virtue of the \textit{Bianchi identities}  $\nabla_a G^{ab} = 0$. But in Geroch–Jang, conservation is simply \textit{assumed}. Moreover, conservation does not entail geodesic motion except under further assumptions—chiefly the absence of internal stresses or interactions.  In realistic systems with pressure, viscosity, or internal structure, conserved matter distributions deviate from geodesics.
This is why the theorem includes the \textit{strengthened dominant energy condition} to ensure that the resulting curve is timelike.

Only in special cases—such as a homogeneous, pressureless dust—does conservation alone imply geodesic motion. There, internal forces vanish by construction, and the dust elements evolve along geodesics even while collectively sourcing the spacetime metric.\footnote{The pressureless dust model represents a continuous, uniform distribution of matter rather than a realistic extended body. In reality, every body described by rigorous physical theories possesses angular momentum, internal stresses, or additional structure (e.g. electromagnetic fields) that typically cause deviations from geodesic motion. More in this in \S\ref{EXTENSION}.}
A paradigmatic instance is provided by the FLRW cosmological model, where non-relativistic matter is modelled as a dust:\footnote{The FLRW framework is not limited to a single dust component. The total cosmic fluid may comprise several components, each with its own equation of state, including non-relativistic matter (e.g., cold dark matter or baryonic matter), radiation, and dark energy. Geodesic motion of fluid elements applies only to the pressureless dust component.  (non-relativistic matter). Relativistic components, such as radiation, possess stress terms in their energy–momentum tensors that lead to deviations from geodesic flow. Individual photons, however, always follow null geodesics in GR, regardless of the fluid description.}
In this case, the energy–momentum tensor takes the form
\begin{equation}
T^{ab} = \rho\, u^a u^b,\label{stressenergydust}
\end{equation}
where \(\rho>0\) is the energy density and \(u^a\) is the four-velocity field of the fluid.
Here, conservation condition entails:
\begin{equation}
\nabla_a T^{ab} = 0 \quad\Longrightarrow\quad u^b \nabla_b u^a = 0,\label{geodesicFLRW}
\end{equation}
which is precisely the geodesic equation.\footnote{Conversely, assuming geodesic motion for the dust fluid implies that $T^{ab}$ is divergence-free and satisfies the strengthened dominant energy condition \eqref{sdec}.} 

But this case is finely tuned: even small departures  from these assumptions destroy the geodesic flow.
In fact, in the slightly more general cases, such as a perfect fluid with \textit{non-zero pressure}, the equation of motion is given by the Euler equation:
$$u^b\nabla_bu^a=-\cfrac{1}{\rho+P}(g^{ab}+u^au^b)\nabla_bP,$$
where $\nabla_bP$ represents the pressure gradient. 
In the most general case of an imperfect, charged fluid, additional terms—bulk and shear viscosity, heat flux, electromagnetic forces, etc.—appear on the right-hand side as non-gravitational forces arising from internal structure, leading to deviations from geodesic motion.

\item[(T3) Extension without extension.] 
Finally, the theorem assumes that each $\underset{\tiny i}{T^{ab}}$ is compactly supported in arbitrarily small neighbourhoods around the curve $\gamma$. This implies that the bodies in question are spatially extended, however slightly. 
But any actual extended body in curved spacetime is subject to tidal effects due to the varying curvature of spacetime across its extent. These effects typically deflect such bodies from geodesic motion.
The point is not that the theorem makes a mistake, but that it deliberately sidesteps this by assuming that the matter fields are smooth, structureless, and free of internal degrees of freedom. In doing so, it constructs extended bodies that are physically empty—possessing no features that define extension. In this sense, the construction undermines its own interpretive basis.

\end{description}

These three tensions expose the central limitation of the Geroch–Jang theorem. It does not derive the geodesic principle from the dynamical content of GR. It assumes conditions under which geodesic motion may be \emph{assigned} to a curve—but those conditions amount to \textit{presupposing} that the motion is geodesic. Thus, the theorem shows only that geodesity can be assigned \textit{to a curve}. 
The background metric supports geodesic motion, but the matter distribution that should instantiate it—namely, a shrinking matter distribution with non-zero $T^{ab}$—cannot coexist with that metric. The motion is not explained; it is imposed.

\paragraph{No claim on Idealisation.}
The Geroch–Jang theorem makes no claim—positive or negative—about idealisation in Norton’s sense. It is not a limit proof: no limiting procedure is invoked, and no attempt is made to construct a limit system that might instantiate a limit property. As such, the theorem does not fall within either of Norton’s two failure modes for idealisation. The result is entirely silent on whether geodesic motion could emerge as the property of a fictitious, idealised system constructed via some asymptotic or structural procedure.

\paragraph{No Approximation.}

What the theorem does reveal, however, is a decisive failure of geodesic motion to serve as an approximation. The curve to which geodesicity is attributed exists in a fixed background spacetime that is not sourced by the matter it contains. This artificial decoupling between the stress–energy tensor and the geometry violates the core structure of general relativity, which requires that matter and geometry co-determine one another through the Einstein field equations. As such, the construction is dynamically incoherent: it violates the theory’s equations. This is an instance of what I have called an \textit{off-shell failure}: there is no regime in which this approximation is valid.

\subsection{The Ehlers-Geroch Theorem}\label{ehlersgeroch}

Whereas the Geroch–Jang theorem approaches the Einstein field equations from the side of the source—constructing a sequence of matter distributions with shrinking support—the Ehlers–Geroch theorem focuses instead on the geometry \citep{Ehlers2004}. . It constructs a sequence of Lorentzian metrics $(\underset{\tiny j}{g_{ab}})_{j\in\mathbb{N}}$ converging for $j\to\infty$ to a fixed background metric $g_{ab}$. The idea is to show that, in the limit, the influence of matter on the geometry becomes negligible, and the limiting curve $\gamma$ exhibits geodesic behaviour.

The theorem is stated as follows:
\begin{thm}
    Let $\gamma:I\to \mathcal{M}$ be a smooth timelike curve in Lorentzian spacetime $(\mathcal{M},g_{ab})$. Suppose that for any sufficiently small closed neighborhood $\mathcal{K}\subset\mathcal{M}$ of $\gamma[I]$ there exists a sequence of smooth Lorentzian metrics $\underset{\tiny j}{g_{ab}}$ defined on $\mathcal{K}$ such that for all points $p\in \mathcal{K}$:
    \begin{enumerate}
        \item for all $j$: $\underset{\tiny j}{G_{ab}}$ has non-vanishing support contained in the interior of $\mathcal{K}$,
        \item for all $j$ and all timelike $\xi^a$: $\underset{\tiny j}{G_{ab}}\xi^a\xi^b\geq 0$ and if $\underset{\tiny j}{G_{ab}}\neq 0$, then $\underset{\tiny j}{g^{bd}}(\underset{\tiny j}{G_{ab}}\xi^a) (\underset{\tiny j}{G_{cd}}\xi^c)>0$,
        \item the $\underset{\tiny j}{g_{ab}}\to g_{ab}$ as metrics in $\mathcal{C}^1(\mathcal{K})$ as $j\to\infty$,
    \end{enumerate}
where $G_{ab}$ is the Einstein curvature tensor determined by $\underset{\tiny j}{g_{ab}}$, then $\gamma[I]$ is the image of a $g_{ab}$-geodesic.
\end{thm}

This result improves upon the Geroch–Jang theorem in one crucial respect: it takes into account the backreaction of the matter source—so long as its effect becomes negligible in the limit. Thus, the stress–energy configurations are not placed in a fixed background but are understood to source the sequence of metrics $\underset{\tiny j}{g_{ab}}$.
The \textit{Ehlers–Geroch particles} (EG-particles), as they are sometimes called, are localised, backreacting bodies whose gravitational influence diminishes in the limit.
That is, the geodesic behaviour of $\gamma$  thus persists \textit{even when backreaction is accounted for}—provided it is made vanishingly small in the limit.

In this way, the Ehlers–Geroch theorem directly addresses the primary limitation of Geroch–Jang (see (T1) above): it ensures that geodesicity is not merely a property of the background metric, but that the limiting curve $\gamma$ asymptotically approaches a geodesic with respect to the metrics $\underset{\tiny j}{g_{ab}}$  generated by the matter itself.
That is, not only does the limiting metric admit $\gamma$ as a geodesic, but the metrics sourced by the EG-particles approximate this behaviour arbitrarily well. The backreaction problem is explicitly controlled.

Yet this improvement comes at a cost. 
\textit{The matter configuration vanishes at the limit}. 
Since the stress-energy tensor associated with the EG-particles must disappear entirely in the limit $j\to\infty$ (and so do each $\underset{\tiny j}{G^{ab}})$, \textit{the limiting spacetime for which the geodesic motion is defined contains no massive body at all}. The geodesic property of $\gamma$ is thus recovered only by excising the body that was supposed to justify it.

This implication is philosophically revealing. 
Geodesic motion as a limit property is preserved only at the cost of erasing its material referent.  The EG-particles approach geodesic motion, only by ceasing to exist. 

\paragraph{No Idealisation.}
From the standpoint of Norton’s framework, this constitutes a failure of the second kind: the limit system exists, but it does not bear the limit property. The Ehlers–Geroch theorem constructs a well-defined limiting system—$(g_{ab}, 0)$—to which the sequence of matter-filled spacetimes converges. But that limit system contains no matter at all. The stress–energy tensor vanishes, and thus the system describes pure vacuum. The geodesic property of $\gamma$ remains, but it is no longer associated with the motion of a massive body. There is no idealised system that instantiates the property in question. In Norton’s terms, the limit property and the limit system disagree: the former requires a body in motion, the latter contains no body at all.

\paragraph{No Approximation.}
One might still ask whether the result justifies geodesic motion as an approximation. But approximation, in Norton’s sense, requires that the property in question—in this case, motion along a geodesic—be an inaccurate but meaningful description of a real target system governed by the full theory. That condition is not met. At each finite stage of the sequence, the body is present, its stress–energy is non-zero, and it backreacts on the geometry. Its motion is not geodesic. In the limit, the body disappears altogether. There is therefore no point in the construction—neither in the sequence nor in the limit—at which a real, dynamically admissible body follows a geodesic, even approximately. 
This is a failure of approximation because the entire approximating sequence fails to represent the target system’s motion at any stage. The sequence drops the referent it aims to describe, yielding a \textit{pathological tracking} in the sense of Norton.

\vspace{0,5cm}
In short: although the Ehlers–Geroch theorem offers a more refined account of how geodesic motion might emerge, it ultimately fails to justify the geodesic principle either as an idealisation or as an approximation. It confirms, rather than refutes, the suspicion that geodesic motion is not the behaviour of any physically admissible system in GR. It is a formal artefact of the geometric framework—well-defined mathematically, but without any referent in the space of dynamically admissible systems.

\subsection{The Einstein-Grommer Proof.}\label{einsteingrommer}

In their 1927 collaboration, \cite{EinsteinGrommer}  explicitly rejected the strategy of representing matter through smooth stress–energy fields.\footnote{As \citet[p.22]{Weatherall2016-WEAIME} notes: \begin{quote} a principal difficulty in trying to derive the geodesic principle as a theorem [that is, as a consequence of dynamics and not as an independent postulate] concerns a kind of ontological mismatch between the geodesic principle and the rest of general relativity: namely, general relativity is a field theory, whereas the geodesic principle is a statement concerning point particles.\rq{}\rq{} (my aside). \end{quote}}
As Tamir notes, they viewed the field-theoretic treatment of matter as conceptually inadequate. 
Einstein in particular considered such representations a \lq{}low-grade\rq{} approximation \citep{einstein1954ideas},  ill-suited to capture the discrete, particle-like character of gravitating bodies.\footnote{The broader metaphysical debate over whether physics should be interpreted in terms of particles or fields remains unresolved, especially in the context of quantum field theory. For contrasting perspectives, see \cite{kuhlmann2002ontological, Glick2016, Jia2022}.} 
Their goal was to eliminate altogether the dualism between field and matter.

This led to what they termed the \lq{}\textit{third way}\rq{}: an attempt to derive the motion of massive bodies directly from the vacuum Einstein equations, without ever introducing a stress–energy tensor.
Rather than model matter by $T^{ab}$, they proposed to \emph{confine} the entire material content of a body to a singular worldline, which is \textit{excised} from the manifold.
The resulting strategy—what Tamir classifies as a \textit{singularity proof}—aims to deduce geodesic motion as a formal consequence of the vacuum Einstein equations. The logic proceeds as follows:

\begin{itemize}
\item A material body is \lq{}confined\rq{} to a \textit{singular}, i.e. one-dimensional, worldline. 
\item The \textit{singular} worldline $\gamma$ is excised from the spacetime manifold $\mathcal{M}$. As a result, the field equations are solved in the domain $\mathcal{M} \setminus  \gamma$, where the EFEs reduce to vacuum form $R_{ab}=0$.
\item A vacuum solution is obtained in the domain $\mathcal{M} \setminus \gamma$, with appropriate boundary conditions imposed near the excised curve $\gamma$. This solution is interpreted as encoding the gravitational field generated by the absent body.
\item The singular worldline $\gamma$  is then reinserted and interpreted as a geodesic of the surrounding vacuum spacetime.
\end{itemize}

Like the Geroch–Jang theorem, this strategy avoids backreaction.
But it does so in a more radical way: not by neglecting the metric response to matter, but by eliminating matter altogether. No stress–energy tensor is introduced. The motion of the body is supposed to emerge from the vacuum spacetime alone.

At first glance, this may seem more consistent than the unjustified test-body regime of Geroch–Jang: no contradiction arises between the presence of stress–energy and the assumption of a fixed background. 
But this consistency is achieved at a profound cost. The singular worldline $\gamma$ lies outside the manifold, so the metric is undefined on $\gamma$, which means the geodesic equation—being a differential relation involving the metric and its derivatives—cannot be formulated at all. 
As Earman observes, \lq{}\lq{}to speak of singularities in $g_{ab}$ as geodesics of the spacetime is to speak in oxymorons\rq{}\rq{} \citep[p.12]{earman1995bangs}.

The conceptual incoherence is thus clear. The very move that removes the need for a stress–energy source also removes the locus of motion.\footnote{This is analogous to the treatment of black hole singularities. See \cite{Curiel2019-CURSAB-2}; see also \cite[p.12]{EarmanJohn1995Bcwa} who observes that \lq{}\lq{}to speak of singularities in gabgab as geodesics of the spacetime is to speak in oxymorons.\rq{}\rq{}.}
There is no longer any physical system to which the geodesic property can be ascribed. What remains is a formal artefact: a curve \textit{a posteriori} interpreted as the trajectory of a body.

\paragraph{No Claim on Idealisation.}
This strategy does not construct a sequence of systems and involves no limiting procedure. It does not define a limit system nor compare it to a limit property. Accordingly, it does not fall within the scope of Norton’s framework for idealisation. Like the Geroch–Jang theorem, the Einstein–Grommer strategy makes no claim—positive or negative—about whether geodesic motion arises as the limit property of an ideal system constructed from a converging sequence of systems. It simply bypasses the question.

\paragraph{No Approximation.}

However, in the case of approximation, the theorem delivers a conclusive verdict. In light of Norton’s framework, the Einstein--Grommer singularity proof is an outright disqualification from approximation. 
In Norton’s framework, approximation requires that the property under consideration—here, geodesic motion—serve as an inaccurate but meaningful description of a real target system. The issue is not whether such a system exists—of course it does: real bodies, with mass and internal structure, are everywhere in GR. 
The failure lies in the fact that the Einstein–Grommer proof  removes the body from the manifold entirely. There is no \lq{}geodesic body\rq{} in the geometry, and no consistent justification why the geodesic motion could approximate the motion of a real target one. The geodesic curve refers to nothing.

This diagnostic closely parallels the failure seen in the Ehlers–Geroch construction: there, too, the geodesic trajectory persists while the material system vanishes in the limit. But the two failures differ in structure. Ehlers–Geroch begins with a sequence of fully dynamical systems—each with well-defined matter and backreaction—and recovers a vacuum geodesic as the limiting case, with the body disappearing in the limit. Einstein–Grommer, by contrast, never introduces any material system at all: the body is excised from the start, and the vacuum equations are solved in its absence. There is no candidate system whose behaviour is even approximately captured by the geodesic.

The Einstein-Grommer construction is a textbook case of what I termed a \textit{pathological tracking}: the geodesic approximation is a formal construction that \textit{tracks nothing}.

However, as the next subsection will make clear, the failure may also be interpreted more strongly as a case of \textit{off-shell failure} of the approximation procedure. For while vacuum geodesics are perfectly consistent with the Einstein field equations, attributing (even if approximately) \textit{a posteriori} the geodesic motion along $\gamma$ to a massive body explicitly violates EFEs. The body can be light, but its stress–energy would need to be supported on a one-dimensional curve. That possibility is explicitly ruled out by the Geroch–Traschen theorem.

\vspace{0,5cm}
In this respect, the Einstein–Grommer strategy offers a particularly stark philosophical lesson. It reveals that geodesic motion in GR cannot even be inaccurately attributed to real bodies, because it is not the motion of any body at all. It is a property of vacuum geometry near a hole in the manifold—a mathematical residue without a referent.
This conclusion reinforces the broader claim advanced throughout this section: geodesic motion in GR is not an approximation in any legitimate sense.

\subsection{The Geroch-Traschen Theorem}\label{gerochtraschen}

The final and most decisive challenge to the geodesic principle as a physically grounded statement comes from a theorem by \cite{Geroch1987} which proves that the EFEs admit no solutions in which the stress–energy is supported on a one-dimensional curve.\footnote{Geroch and Traschen introduce a class of metrics now known as\textit{ GT-regular metrics}: Lorentzian metrics whose components and inverses are locally integrable, and whose weak derivatives are locally square-integrable. These regularity conditions ensure that the stress-energy tensor, which involves second derivatives of the metric, is well-defined as a \textit{tensorial distribution}. This allows the use of objects like the \textit{Dirac delta} to model highly concentrated mass-energy sources. The hope behind singular derivations of the geodesic principle is that the stress–energy of a massive body could be represented by such a distribution supported on a timelike curve. The Geroch–Traschen theorem rules this out: no such distributional source is compatible with the EFEs under GT-regularity.} 

This result is not a limit proof; it does not construct a sequence of systems. 
Instead, it plays a \lq{}blocking role\rq{}: it shows that the very \textit{endpoint} of any limit strategy involving point-particles is \textit{inadmissible}.
As such, it blocks also the Einstein–Grommer strategy discussed in the previous subsection, and any other attempt to model massive point particles via distributional stress–energy supported on a singular curve. In fact, although the Einstein–Grommer proof does not involve a limit construction, it nevertheless assumes that a massive point source—albeit excised—can be meaningfully associated with a geodesic.

In this sense, the Geroch–Traschen theorem belongs to the same philosophical class as Einstein–Grommer: it is a \textit{singularity result}, in that it diagnoses the incompatibility between GR and a class of singular source models, that is, one-dimensional distributional sources.
But unlike Einstein–Grommer, which relies on excising the singularity from the manifold and retroactively assigning geodesicity, Geroch–Traschen proves a more general \textit{no-go theorem}.

\paragraph{No Idealisation.}
In Norton’s framework, this is a paradigmatic case of a failed idealisation of the first kind: the \emph{limit system does not exist}. To be precise, it is an even stronger case: the limit system \textit{cannot} exist.
Even if the geodesic property is well-defined as a mathematical limit, the limit system to which it is supposed to apply cannot be constructed within the theory. 
The nonlinearity of the Einstein field equations plays a decisive role here. In linear theories, or in the linearised approximation to GR, such singular constructions can often be handled safely, as we will stress in \S\ref{perturbed}.T But in full GR, highly concentrated energy–momentum distributions generate curvature singularities incompatible with the field equations. The theory resists the representation of one-dimensional mass distributions.

The upshot is clear. No dynamically admissible solution can represent a massive point particle moving along a geodesic. As such, no sequence of extended stress–energy configurations can converge to a geodesically moving point-mass.
The limit object simply lies outside the space of admissible solutions. GR, in its full non-linear form, cannot accommodate one-dimensional distributions of mass-energy. There is no admissible way to concentrate stress–energy onto a worldline without violating the field equations.

This renders the familiar physical reasoning—according to which geodesic motion emerges in the limit as small bodies shrink and internal structure vanishes—deeply problematic. 
If the field equations rule out the limit system, then the limit property is empty. Geodesic motion is not the behaviour of an ideal system; it is the residue of a failed construction.
It is a limit property with no admissible limit system—a purely formal artefact, disconnected from the representational structure of the theory.

\paragraph{No Approximation.}
At first glance, the Geroch–Traschen theorem might seem irrelevant to approximation claims, since geodesic motion concerns test bodies—systems that do not source the metric—whereas the theorem blocks massive point particles viewed as singular \textit{sources}.
But this appearance is misleading.
To approximate the motion of a real, extended, backreacting body by a test body geodesic motion, we must construct a procedure according to which the target body becomes point-like (that is supported on a 1D curve), becomes very light (so that backreaction becomes negligible) and yet retains enough mass to follow a \textit{timelike} trajectory, all while existing within the representational structure of the theory.

The Geroch–Traschen theorem blocks any such procedure. It prohibits any attempt to represent a target body as a massive body whose motion is both geodesic and concentrated on a worldline. Therefore it precludes the geodesic property from serving as a physically grounded approximation.
The failure is not because the endpoint of the process (geodesic motion) is inadmissible, but because the procedure itself—concentrating a backreacting body to a 1D massive source—cannot be carried out in a way that remains consistent with the EFEs.
The familiar practice of modelling such bodies as test particles moving on geodesics is not an approximation in the technical sense, it is a formal artefact disconnected from the theory’s space of physically admissible behaviours.
As should now be clear, this constitutes an \textit{off-shell failure} of approximation.
\vspace{0.5cm}

This completes the dismantling of the geodesic principle as a candidate law of motion. In every attempted derivation—whether by singular limit, or by excision—the principle either presupposes its conclusion, loses its referent, or fails to be borne by any admissible system. The following section turns to what geodesic motion was thought to approximate: the motion of extended, structured bodies.

Table \ref{tab:theorems} below summarises the role of the theorems analysed in challenging the validity of the geodesic principle as a possible approximation or idealisation to natural motion of bodies.
\clearpage

\begin{sidewaystable}[ht]
\centering
\renewcommand{\arraystretch}{1.6}
\begin{tabular}{|>{\raggedright\arraybackslash}p{3.5cm}
                |>{\raggedright\arraybackslash}p{5.2cm}
                |>{\raggedright\arraybackslash}p{6.1cm}
                |>{\raggedright\arraybackslash}p{6.1cm}|}
\hline
\textbf{Theorem / Strategy} 
& \textbf{Purpose and Method} 
& \textbf{Failure as Approximation} 
& \textbf{Failure as Idealisation} \\
\hline

\textbf{Geroch–Jang} 
& Attempt to derive geodesic principle from EFEs by showing that any smooth, symmetric, conserved stress-energy tensor $T^{ab}$ with compact support in a neighbourhood of a curve $\gamma$ implies that $\gamma$ is a geodesic.& Geodesicity is assigned in a fixed background  to bodies (GJ-particles) possessing non-zero $T^{ab}$ is present. The construction artificially decouples matter and geometry. \newline
\textbf{Failure type: Off-shell}.& The theorem does not involve a limiting procedure. No limit system is constructed or implied. \newline
\textbf{No claim on idealisation}.\\
\hline

\textbf{Ehlers–Geroch} 
& Attempt to derive geodesic principle from EFEs including backreaction. Construct a sequence of spacetimes $(\mathcal{M},\underset{\tiny j}{g_{ab}}, \underset{\tiny j}{T^{ab}})$ sourced by localised matter distributions, such that the sequence converges  to a vacuum spacetime $(\mathcal{M},g_{ab}, 0)$ admitting $\gamma$ as a geodesic.& No stage of the construction—neither in the sequence nor the limit—approximates the motion of a real body. \newline
\textbf{Failure type: Pathological tracking}.& The limit system $(g_{ab}, 0)$ exists but does not instantiate the limit property (geodesic motion of a body). \newline
\textbf{Type II failure}: limit property and limit system disagree.\\
\hline

\textbf{Einstein–Grommer} 
& Attempt to derive geodesic principle from EFEs by solving the vacuum EFEs on $\mathcal{M} \setminus \gamma$ and \textit{a posteriori} interpreting the excised geodesic $\gamma$ as the body’s path. No $T^{ab}$ is introduced.& There is no body in the manifold to approximate; $\gamma$ refers to nothing physical. The approximation tracks no system at all. \newline
\textbf{Failure type: Pathological tracking (and possibly Off-shell, cf. Geroch–Traschen)}.& The proof does not involve a limiting procedure. No limit system is constructed or implied. \newline
\textbf{No claim on idealisation}.\\
\hline

\textbf{Geroch–Traschen} 
& Prove that the EFEs (under GT-regularity) do not admit stress–energy distributions supported on 1D curves. It is a \textit{no-go theorem}.& The theorem blocks any procedure that concentrate a target body on a 1D curve. The procedure takes us \textit{off-shell}.\newline
\textbf{Failure type: Off-shell}.& The theorem blocks the construction of the limit system: a stress–energy supported on a 1D curve. No limit system exists to bear the geodesic property. \newline
\textbf{Type I failure}: no limit system exists.\\
\hline

\end{tabular}
\caption{Four canonical strategies to derive the geodesic principle from GR and their role within the epistemic framework developed in this paper to challenge the validity of the geodesic principle as a possible approximation or idealisation to natural motion of bodies.}
\label{tab:theorems}
\end{sidewaystable}

\clearpage

\section{Extended Test Bodies}\label{EXTENSION}

The systems considered in this section are test bodies that retain internal structure and finite spatial extension, but are assumed not to generate curvature—that is, they do not backreact on the spacetime geometry. They represent an intermediate stage in the analysis of free motion: complex enough to register systematic departures from geodesic behaviour, yet simple enough to avoid the dynamical complications introduced by backreaction. As such, they provide a natural testing ground for assessing the limits of the geodesic principle.

The conclusion of my analysis will be that if geodesic motion is taken to define inertial motion, then spatially extended bodies—even in the test-body regime—do not move inertially. Their internal structure interacts with background curvature in ways that deflect them from geodesic paths. This undermines any attempt to ground a physically meaningful notion of natural motion in the geodesic motion.
Instead, the actual trajectories of extended test bodies must be described using more refined frameworks, which remain consistent with the Einstein field equations, even though the bodies themselves are not sources of curvature.

This raises a more precise question: how should one model the motion of such extended test bodies, whose internal structure induces systematic deviations from geodesic motion, even in the absence of backreaction? Two complementary formalisms provide leading-order approximations within the broader hierarchy of natural motion:

(i) The \textit{Mathisson–Papapetrou–Dixon (MPD) equations}, which govern the motion of the body’s centre of mass, taking into account spin and higher multipole moments. These equations are derived from a multipole expansion of the body's stress–energy tensor and encode how internal structure couples to background curvature (\S\ref{MPD}). In this context, the stress–energy tensor $T^{ab}$  is not interpreted as a source of curvature, but as a formal device used to represent internal structure—just as it was in the derivation of the geodesic principle via the Geroch–Jang theorem. However, the philosophical posture is crucially different. In the MPD framework, the equations of motion are not claimed to follow from EFEs. They provide a well-defined approximate model to isolate and analyse the effect of \textit{internal structure alone}.  In contrast, the Geroch–Jang theorem purports to derive geodesic motion from constraints on matter fields satisfying the EFEs and yet the derivation ignores backreaction. This makes the derivation internally inconsistent, if it is interpreted as giving rise to physically valid motion.

(ii) The \textit{geodesic deviation equation}, which captures the relative acceleration between neighbouring geodesics in a congruence that models the extended body. This framework describes the influence of curvature gradients across the body's \textit{interior}, including tidal effects and structural deformations (\S\ref{deviation} ).

The remainder of this section is devoted to analysing these two frameworks. Together, they illustrate how natural motion can be reconceptualised beyond the geodesic artefact. 
In doing so, they begin to reveal the layered structure that replaces the geodesic principle in GR.

\subsection{A First Step Beyond Geodesics: Spin and Torsion}\label{MPD}

Two instructive examples illustrate that gravitational motion of test bodies depart systematically from geodesic behaviour: the presence of \textit{spin} and the influence of \textit{torsion}.
 Each reflects a distinct mechanism by which internal structure or geometry determine the path of a body, underscoring the inadequacy of the geodesic principle as a general description of free fall.

\textit{Spin–curvature coupling} represent the leading-order contributions to the motion of spatially extended test bodies with internal angular momentum. At dipole order, this interaction captures the effects of spin, while higher multipole terms account for more detailed aspects of the internal configuration, such as quadrupole moments. These cases differ in detail but converge in implication: geodesic motion is not even the starting point of the approximation hierarchy of natural motion, since the motion of extended test bodies (e.g. governed by the MPD equations) is a \textit{first physically meaningful approximation} within GR.

\textit{Torsion}, presents a complementary case. Rather than modifying the motion through internal structure, it alters the very geometry that defines free fall. In theories such as Einstein–Cartan gravity, torsion contributes to the affine connection and thereby modifies the paths of structureless bodies. While such models extend beyond standard GR, they offer a powerful illustration of how small alterations to the geometrical structure can yield physically consistent—but non-geodesic—forms of free fall.

\subsubsection{Spin.}

\begin{description}
\item[Dipole Expansion.] Consider test bodies with internal angular momentum—so-called spinning bodies—moving in a curved but torsion-free spacetime. Even in the absence of external forces, such bodies do not follow geodesics. Their motion is governed by the Mathisson–Papapetrou–Dixon (MPD) equations, which encode how spin couples to background curvature and deflects the trajectory from geodesic motion \citep{Papa1951,dixon1970dynamics}:
\begin{equation}
    \frac{D p^a}{d\tau} = -\frac{1}{2}\,R^a{}_{bcd}\,u^b\,S^{cd},   \quad\quad \frac{D S^{ab}}{d\tau} = 2\,p^{[a}\,u^{b]},
\end{equation}
where $\frac{D}{d\tau}$ is the usual covariant derivative along the representative worldline of the centre of mass, parametrised by proper time $\tau$; $p^a$ is the 4-momentum; $u^a$ is its four-velocity; and $S^{ab}$ as the antisymmetric spin tensor. 

The spin tensor represents the \textit{dipole} moment of the body’s angular momentum, defined relative to a representative worldline associated with the centre of mass.
While the body is modelled as spatially extended, these quantities are defined along this representative trajectory, which serves to represent the entire body.
The body's internal structure is instead encoded via multipole moments, which are projected onto the representative worldline.\footnote{The system must be closed by imposing a supplementary condition, such as the Tulczyjew–Dixon constraint ($S^{ab}p_b=0$) \textit{or} the Mathisson–Pirani constraint ($S^{ab}u_b=0$) to determine how to define the centre of mass worldline. See \cite{dixon1970dynamics}.} 
In particular, $S^{ab}$ should be interpreted as an \textit{effective} spin tensor: it encodes the dipole moment of the body's angular momentum about the centre-of-mass worldline, without resolving the full internal dynamics or non-rigid rotation.
In the MPD formalism, this marks the first physically admissible level of structural complexity in the motion of test bodies—beyond which further contributions, such as quadrupole terms, refine the description.

\end{description}

\begin{description}
\item[Quadrupole Expansion.]
At quadrupole order, the MPD formalism incorporates contributions that depend on how curvature varies across the finite spatial extent of the body. Although the body is still represented by a single worldline (that of the centre of mass), the quadrupole moment captures how its internal structure couples to curvature gradients in the background spacetime. These effects arise because different regions of the body interact with slightly different ambient curvature, and their net influence modifies the motion of the centre of mass. In this sense, the formalism reflects spatial extension more finely than at dipole order, where only the integrated angular momentum contributes.

The quadrupole tensor $J^{abcd}$,  which satisfies the same algebraic symmetries as the Riemann tensor,
\begin{equation}
    J^{abcd}=J^{[ab][cd]}=J^{cdab}, \quad J^{[abc]d}=0,
\end{equation}
encodes this quadrupole structure. No particular assumption is made here about the origin of $J^{abcd}$: it may include contributions from the mass distribution, internal stresses, or other structural features \citealp[\S6]{dixon1970dynamics}. 
 
The MPD equations at this order read:
\begin{equation}
    \frac{D p^a}{d\tau} = -\frac{1}{2}\,R^a{}_{bcd}\,u^b\,S^{cd}-\frac{1}{6}J^{bcde}\nabla^a R^{bcde}.
\end{equation}
The second term represent the curvature gradients-quadrupole coupling.

This term remains consistent with the test-body approximation: TabTab enters kinematically to define structure, while the background metric remains fixed. Quadrupole terms thus mark the upper bound of internal complexity compatible with a non-backreacting body. Beyond this, further contributions typically require the inclusion of backreaction.\footnote{The MPD formalism typically becomes intractable beyond quadrupole order and generally require incorporating backreaction.}

At first glance, such quadrupole effects might appear \lq{}tidal\rq{} in nature, since both arise from curvature gradients
But their domains differ. In the MPD formalism, quadrupole terms determine how the body's internal structure influences the motion of its centre of mass, projected onto a single worldline. They govern the \textit{overall acceleration} of the body’s centre of mass.
They are sometimes called \textit{tidal effects}, but this can cause confusion, because in GR tidal effects typically refer to how curvature gradients induce\textit{ relative acceleration} of infinitesimally nearby worldlines within an extended configuration. These approaches serve complementary functions in the broader theory of natural motion.

While both arise from curvature gradients, the MPD quadrupole terms govern the acceleration of the body's \textit{centre of mass} in response to curvature, whereas geodesic deviation captures the \textit{internal relative acceleration} within an extended configuration. Their domains differ, but both represent leading-order curvature-sensitive contributions consistent with the test-body regime.
The formal and conceptual relationship between these two approaches is taken up in detail in \S\ref{deviation}, where their complementarity is clarified.

\end{description}

\subsubsection{Torsion.}\label{torsion} 

A further perspective on the limitations of geodesic motion arises not from the internal structure of bodies, but from the geometry of spacetime itself. 
In gravitational theories that extend GR to include torsion—such as Einstein–Cartan theory—the identification of inertial motion with Levi–Civita geodesics fails in a more structural way \citep{Cartan1922,Penrose1983,Hehl1995}. 
These theories employ a general affine connection that need not be symmetric; its antisymmetric part defines the \textit{torsion tensor}, which modifies both parallel transport and the equations of motion for test bodies.

This generalisation has immediate implications. 
In spacetimes with torsion, the two standard characterisations of geodesics diverge: \textit{extremals} of the spacetime interval no longer coincide with \textit{self-parallel curves}.

The \textit{Levi–Civita connection}, which remains metric-compatible and symmetric, defines the extremals.\footnote{The metric compatibility condition is $\nabla_cg_{ab}=0$ and is also referred to as the \textit{Ricci Theorem} in tensor analysis. The Levi-Civita connection coefficients (given a basis), also called \textit{Christoffell's symbols}, are given by $\Gamma^a_{bc} = \frac{1}{2} \, g^{ad} \left( \partial_b g_{dc} + \partial_c g_{bd} - \partial_d g_{bc} \right)$. These define extremal curves but do not generalise to torsional spacetimes.}

By contrast, the \textit{general} \textit{affine connection} that governs parallel transport is defined independently of the metric and need not be symmetric.\footnote{An affine connection can be introduced via the covariant derivative acting on basis vectors along tangent directions. No metric structure is required for its definition. Christoﬀell symbols, by contrast, are defined from the metric structure of the manifold.}
Its antisymmetric part defines the torsion tensor. In the absence of torsion, the connection reduces to the Levi–Civita form.
As a result, is spacetime with torsion, extremal curves and self-parallels no longer coincide.

Instead, free motion is governed by dynamical laws that account for \textit{torsion–matter coupling}. 
Crucially, torsion couples only to \textit{spin}, not to mass-energy alone. 
This implies that a scalar or spinless (hence, point-like) test body feel no influence from torsion at all, while a spinning body—those with non-vanishing antisymmetric dipole moment—experience additional forces and torques due to torsion.

To model these effects, one must appeal to a generalised version of the MPD equations at the dipole order. 
In this framework, the motion of spinning test bodies is governed jointly by curvature and torsion, with the contortion tensor mediating the spin–torsion interaction.

The generalised MPD equations in a torsionful spacetime take the form \citep{Hehl1995,Blagojevi2011}:

\begin{equation}
    \frac{D p^a}{d\tau} = -\frac{1}{2}\,\tilde{R}^a{}_{bcd}\,u^b\,S^{cd} - K^a{}_{cd}\,\frac{D S^{cd}}{d\tau}, \quad \quad \frac{D S^{ab}}{d\tau} = 2\,p^{[a}\,u^{b]} + 2\,K^{[a}{}_{cd}\,S^{b]d}\,u^c,
\end{equation}

where $\tilde{R}^a{}_{bcd}$ is the \textit{general} curvature tensor associated with the torsionful connection, and $K^a{}_{cd}$ is the \textit{contortion tensor}. 
The torsion tensor is defined by $\Gamma^a{}_{[cd]}=\frac{1}{2}\bigl(\Gamma^a{}_{cd} - \Gamma^a{}_{dc}\bigr)$ (with $\Gamma^a_{cd}$ the usual symmetric Levi-Civita connection), while the contortion is related to the torsion via $K^a{}_{cd} = \frac{1}{2} \left( \Gamma^a{}_{[cd]} - \Gamma^b{}_{[ca]} g_{d b} + \Gamma^b{}_{[d a]} g_{cb} \right)$, with square brackets indicating antisymmetrisation over the enclosed indices.

These equations illustrate that free motion is not determined solely by the intrinsic properties of bodies, such as mass or spin, but arises from their dynamical interaction with the geometric structure of spacetime.
In torsional spacetimes, there is no universal equation of inertial motion. 

For spinning test bodies, the generalised MPD equations—incorporating curvature and torsion via the contortion tensor—constitute the most basic dynamical law of free fall. No appeal to geodesics, whether extremal or autoparallel, is relevant in this regime. In fact,  as  \cite{Weatherall2016-WEAIME} emphasises, in the presence of torsion, extremals and self-paralleles no longer coincide, and neither yields a general criterion for inertial motion.

In contrast, spinless bodies—such as scalar particles—do not couple to torsion and are typically assumed to follow the extremals of the Levi–Civita connection. 

The appropriate dynamical equation of free fall must be identified case by case, depending on the body’s internal structure and the geometry of the theory. This reinforces the broader methodological point: what qualifies as \lq{}inertial\rq{} motion varies with the representational framework—there’s no unique geometrical account.


\subsection{Geodesic Deviation: Tidal Effects}\label{deviation}

The geodesic deviation formalism models how spacetime curvature induces relative acceleration between nearby geodesics. Within the test-body regime, it provides a natural way to account for \textit{tidal deformation} in spatially extended test bodies, that experience differential gravitational influence across their structure.

In this setting, the extended body is modelled as a congruence of infinitesimally nearby, non-intersecting timelike geodesics: each follows its own geodesic equation, but the congruence as a whole is subject to curvature-induced distortion.\footnote{In relativity, perfect rigidity is physically inadmissible, as it implies superluminal propagation of internal forces. Special relativity allows only \textit{Born rigidity}, a much more restrictive condition requiring no deformation in the body's own instantaneous rest frame.  In curved spacetime, even Born rigidity is typically unsustainable due to the presence of tidal effects. See \cite{Giulini2006,Ziyang} for discussion.}

This formalism complements the MPD framework of §\ref{MPD}. While the MPD equations govern the motion of the centre of mass, geodesic deviation characterises the internal evolution of an extended body, capturing effects due to spatial extension without requiring multipole expansions or backreaction.

Importantly, although the formalism of geodesic deviation is built upon a congruence of geodesics, it does \textit{not} rely on interpreting individual geodesics as physically meaningful trajectories. Rather, it captures the leading-order internal deformation of an extended test body in a fixed background. In particular, geodesic deviation is expressed relative to a reference geodesic, which serves as a benchmark for detecting tidal effects. This might seem to reintroduce a privileged role for geodesic motion. But that role is merely \textit{methodological}: the geodesic functions as a \textit{counterfactual scaffold}---a trajectory the body would follow in the absence of curvature gradients. It corresponds to no real or idealised body, and carries no ontological commitment.
In this sense, the formalism avoids the ontological pitfalls of attributing geodesic motion to possible bodies, and remains valid as a physically meaningful approximation within the test-body regime.

Formally, let
\begin{equation}
\gamma: I \times J \to \mathcal{M},
\end{equation}
be a smooth map from the product of real intervals $I$ and $J$ into a differentiable manifold $\mathcal{M}$, where $\tau\in I$ parametrises proper time along each geodesic, and  \(s \in J\) labels individual geodesics within the congruence.
For fixed $s$, the map  $\gamma_s(\tau) := \gamma(\tau, s)$ defines a timelike geodesic in \(\mathcal{M}\). Thus, the image of $\gamma$  defines a smooth \textit{congruence} of neighbouring geodesics.\footnote{Some texts choose to denote the image of  \( \gamma \) as  \( x^a(\tau,s) \) where the superscript \( a \) does not denote a vector in the sense of $x^a$ being an element of a tangent space. Rather,  \( x^a(\tau,s) \) represents the abstract \lq{}position\rq{} of a point on the geodesic of $\mathcal{M}$.}

Associated with this congruence are two key vector fields:
\begin{itemize}
    \item The \textbf{tangent vector field} 
    \begin{equation}
    u^a = \left(\frac{\partial \gamma}{\partial \tau}\right)^a,
    \end{equation}
    representing the four-velocity along each geodesic;\footnote{Technically, $\frac{\partial \gamma}{\partial \tau}$ is the \textit{pushforward} $d\gamma(\frac{\partial}{\partial\tau})$ induced by $\gamma$  which maps the basis vector \(\frac{\partial}{\partial \tau}\) in the parameter space  \( I \times J \) to vectors in the tangent space  \( T_{\gamma(\tau,s)}\mathcal{M} \) at the point  \( \gamma(\tau,s)=x^a(\tau,s) \).}
    \item The \textbf{deviation vector field} 
    \begin{equation}
    \xi^a = \left(\frac{\partial \gamma}{\partial s}\right)^a,
    \end{equation}
    which connects points on infinitesimally adjacent geodesics at equal values of $\tau$.
\end{itemize}

The evolution of $\xi^a$ is governed by the \textit{geodesic deviation equation} (or Jacobi equation) \citep{Wilkins_Riemannian}:
\begin{equation}
\nabla_u \nabla_u \xi = - R^a{}_{bcd}\, u^b\, \xi^c\, u^d,\label{geodesicdeveq}
\end{equation}
where \(\nabla_u:=u^a\nabla_a\) denotes the covariant derivative along the vector field \(u^a\), and \(R^a_{bcd}\) is the Riemann curvature tensor acting on the pair \((u^a,\xi^a)\).\footnote{In coordinate-dependent formulation, one typically writes the geodesics as \(x^\mu(\tau,s)\), with $u^\mu = \frac{dx^\mu}{d\tau}$ and $\quad \xi^\mu = \frac{\partial x^\mu}{\partial s}.$
The geodesic deviation equation thentakes the form: 
\[
\frac{D^2 \xi^\mu}{D\tau^2} = -R^\mu_{\; \nu\rho\sigma}\, u^\nu\, \xi^\rho\, u^\sigma,
\]
with \(\frac{D}{D\tau} = u^\mu \nabla_\mu\).}

This equation describes how curvature induces \textit{relative acceleration} between nearby geodesics.
Consider two point-like test bodies freely falling along neighbouring geodesics, each carrying an infinitesimal, massless accelerometer. 
Each particle has zero proper acceleration individually, but they exhibit a relative acceleration  $A^\mu:=\frac{D^2 \xi^\mu}{D\tau^2}$, which is determined by the Riemann curvature tensor via the geodesic deviation equation \eqref{geodesicdeveq}.
Such relative acceleration, which is a \textit{tidal effect}, is frame-independent: it cannot be removed by a change of frame—it is an \textit{intrinsic} consequence of spacetime curvature.\footnote{As discussed in \S\ref{EQUIVALENCE}, the equivalence principle guarantees that at any point, one can construct a locally inertial frame in which the Levi–Civita components vanish and the metric is the flat Minkowski space. Conversely, in flat spacetime, one may introduce an accelerating frame that \textit{mimics} the presence of a gravitational field. However, to distinguish genuine gravitational effects from mere frame artefacts, one must examine the \textit{relative} motion of free-falling bodies. It is the behaviour of nearby geodesics—and their deviation under curvature—that reveals the true geometric structure of spacetime.}

More broadly, geodesic deviation reveals the limits of  local flatness.
Since Riemann curvature tensor tensor cannot be eliminated even at a point or along a geodesic, the tidal effects it generates persist even in locally inertial frames \citep{Brown2016}. 
Thus, the popular slogan that \lq{}local validity of special relativity\rq{} must be interpreted with care: it holds only in an infinitesimal neighbourhood around a point, and even there, \textit{only to first order} in curvature.

Tidal effects could be neglected only by considering regions so small that nearby geodesics remain arbitrarily close.
In such cases, in a Fermi local inertial frame, the geodesic deviation vector  $\xi^a$ may be assigned arbitrarily small components $\xi^I\ll\mathcal{O}(1)$. 
Then, even in a region with non-zero curvature, the relative acceleration becomes negligible:
\begin{equation}
\frac{D^2 \xi^I}{d\tau^2} = -R^I_{\; JKL}\, u^J\, \xi^K\, u^L \ll\mathcal{O}(1).
\end{equation}
Rigorously, only in the limit of $\xi^I \to 0$ do tidal effects vanish completely—an abstraction corresponding to the body’s extension shrinking to a mathematical point.
This abstraction eliminates all curvature-induced relative motion and is often invoked to justify the equivalence principle. But, as noted in §\ref{THEOREMS}, it presupposes an object that cannot be realised within GR.

The geodesic deviation formalism is also valuable to assess the status of geodesic motion. When $|\xi^a|$ lies below the threshold of experimental resolution, a freely falling body is often said to \lq{}follow a geodesic\rq{}. 
But this is a heuristic shorthand, not a physically grounded approximation. Its role can be at most methodological, not representational.
This brings us back to the philosophical core of the paper: both geodesic motion and the equivalence principle are best understood as \textit{formal constructs}, expressive of the geometry of the Levi–Civita connection, but without physical referents—neither real nor ideal—and therefore excluded from the hierarchy of legitimate approximations that constitutes natural motion.

By contrast, the geodesic deviation formalism captures a genuine physical approximation: it offers a curvature-sensitive account of relative acceleration in extended test bodies.  It complements the MPD formalism by addressing not the centre-of-mass trajectory, but the internal structure of the body as modelled by a congruence of free-falling constituents. Both formalisms remain dynamically consistent within the bounds of GR.

Geodesic deviation also offers a precise and geometrically grounded account of how curvature manifests is \textit{operationally} detectable via tidal effects.
No actual measurement is local in this strict mathematical sense: every experimental apparatus occupies a finite region of spacetime.
For instance, the formalism underpins the interpretation of gravitational wave experiments. Interferometric detectors like \citep{LIGO} function by monitoring the varying separation between freely suspended mirrors, treated as point-like test bodies in free fall. The oscillatory strain pattern induced by a passing gravitational wave corresponds precisely to the relative acceleration predicted by geodesic deviation, and thus directly encodes curvature information (more precisely, components of the Weyl tensor in transverse traceless gauge).\footnote{LIGO mirrors are suspended in vacuum by multi-stage pendula, designed to isolate them from terrestrial vibrations and non-gravitational forces. Over the timescales and amplitudes relevant for gravitational wave detection, they realise the test-body regime: backreaction is negligible, and tidal effects dominate. While they are often said to \lq{}approximate free fall\rq{}, this should not be taken to imply an actual motion along geodesics—which, as argued in §\ref{THEOREMS}, corresponds to no physically admissible configuration. Rather, their mutual separation encodes the relative acceleration predicted by geodesic deviation.}

\paragraph{The Kinematical Picture.} 

This analysis can be extended kinematically, in the sense that one can describes \textit{how} nearby observers move relative to one another, not \textit{why} they move that way.
The covariant derivative of the velocity field  $u^a$ can be decomposed into three parts:
\begin{equation}
\nabla_a u_b = \omega_{ab} + \sigma_{ab} + \frac{1}{3} \theta h_{ab},\label{kinematic}
\end{equation}
where $h_{ab}=g_{ab}+u_a u_b$ is the \textit{projection tensor} onto spatial hypersurfaces orthogonal to $u^a$ and $\nabla_a u_b=k_{ab}$  is sometimes called the \textit{extrinsic curvature}.
The terms on the r.h.s. represent:
\begin{itemize}
\item $\theta:$ the \textit{expansion scalar}, indicating isotropic divergence or convergence of the volume element defined by the congruence;
\item $\sigma_{ab}:$ the \textit{shear tensor}, encoding anisotropic shape deformation without volume change;
\item $\omega_{ab}:$ the \textit{vorticity tensor} , characterising the twist of neighbouring geodesics.\footnote{Vorticity corresponds to Born-rigid,  distance-preserving rotation only in the absence of expansion and shear. More generally, it signals the failure of hypersurface orthogonality.} 
\end{itemize}

This decomposition provides the \lq{}instantaneous\rq{} kinematical state of the congruence. 
Nonetheless, these quantities also satisfy well-defined evolution equations, derived from the Ricci identity and EFEs. These equations reveal the dynamical role of spacetime curvature in shaping the congruence's behaviour over proper time.\footnote{The \textit{Ricci identity} expresses the non-commutativity of covariant derivatives on a vector field: $\nabla_a\nabla_bu^c-\nabla_b\nabla_au^c=R^c_{dab u^d}$, where $R^c_{dab u^d}$ is the Riemann curvature tensor. When applied to the congruence's velocity field $u^a$, this identity generates evolution equations for the expansion, shear, and vorticity tensors.}

Each of the three kinematical quantities satisfies its own evolution equation. The set of the equations are often called the \textit{Raychaudhuri equations} \citep{Hensh2021}.
\begin{enumerate}
\item The equation governing the evolution of $\theta$ is:
\begin{equation}
\frac{d\theta}{d\tau} = -\frac{1}{3}\theta^2 - \sigma^{ab}\sigma_{ab} + \omega^{ab}\omega_{ab} - R_{ab} u^a u^b.
\end{equation}

Here, expansion is damped by shear, enhanced by vorticity, and sourced by the Ricci tensor $R_{ab}$, which encodes the local matter content.
\item The equation governing the evolution of $\sigma_{ab}$ is:
\begin{equation}
\frac{D\sigma_{ab}}{d\tau} = -\frac{2}{3}\theta\,\sigma_{ab} - \sigma_{ac}\sigma^c_{\;b} - \omega_{ac}\omega^c_{\;b} + E_{ab}.
\end{equation}
Here, $E_{ab}$ denotes the so-called \textit{electric part} of the Weyl tensor, which encodes the tidal component of curvature not determined by local matter.\footnote{The Weyl tensor $C_{abcd}$ encodes the trace-free part of the Riemann tensor, representing the tidal and radiative degrees of freedom of the gravitational field that are not locally determined by matter. Relative to a unit timelike vector field $u^a$, it can be decomposed into two symmetric, trace-free, spatial tensors: the \textit{electric part} $E_{ab}:=C_{abcd}u^cu^d$, which governs tidal deformation; and the \textit{magnetic part} $H_{ab}:=\frac{1}{2}\epsilon_{acde}C^{de}_{\;\;bf}u^cu^f$, which encodes frame-dragging and gravitational-wave-like effects. In the evolution of a geodesic congruence, only the electric part explicitly in the shear propagation equation.} This equation shows that shear is dynamically sourced by the tidal component of the curvature and is coupled to both expansion and vorticity.
\item The equation governing the evolution of $\omega_{ab}$ is:
\begin{equation}
\frac{D \omega_{ab}}{d\tau} = -\frac{2}{3} \theta\, \omega_{ab} - 2\, \sigma_{[a}{}^{c} \omega_{b]c}.
\end{equation}
Notably, vorticity evolves independently of curvature and is sourced purely by shear and expansion and remains dynamically decoupled from curvature unless additional structures, such as torsion, are present.
\end{enumerate}

Together, these three evolution equations provide a full dynamical characterisation of the congruence’s local behaviour under gravity. They describe how the spacetime curvature—not just from local matter via the Ricci tensor, but also from tidal structure via the Weyl tensor—governs the distortion, rotation, and divergence of extended test bodies modelled as congruences of free-falling worldlines.

In highly symmetric spacetimes like FLRW cosmology, homogeneity and isotropy impose $\omega_{ab}=\sigma_{ab}=0$, while the expansion scalar is proportional to the Hubble parameter  $\theta \propto H$, encoding the rate of expansion of the Universe \citep{weinberg1972gravitation}.\footnote{Furthermore, in FLRW spacetime, the \textit{synchronous frame} selects $u^\mu=(1,0,0,0)$ and $g_{00}=1,g_{0i}=0$. Consequently, the projection tensor coincides with spatial metric $h_{ij}$ adapted to constant-cosmic time hypersurfaces. Deviations from perfect FLRW symmetry—such as anisotropies or gravitational waves—reintroduce shear and vorticity. However, for \textit{hypersurface orthogonal congruence of timelike geodesics} satisfying $u_{[a}\nabla_b u_{c]}=0$ , the vorticity tensor necessarily vanishes. This condition is automatically satisfied in \textit{globally hyperbolic spacetimes}.}
But in perturbed or anisotropic settings—such as Bianchi models, gravitational waves, or inhomogeneous collapse—shear and vorticity re-emerge as key signatures of deviation from geodesic uniformity.

Taken together, the MPD and geodesic deviation formalisms demonstrate that even in the absence of backreaction, spatial extension and internal structure lead to systematic departures from geodesic motion. These frameworks constitute the first physically meaningful levels in the approximation hierarchy of natural motion. While they retain the assumption of a fixed background geometry, they already make clear that geodesics do not describe the trajectories of realistic bodies.

The next section turns, even if not exhaustively from a technical-formal point of view (this will be a task for future work), to the case of backreacting systems. 
There, the concept of natural motion must be extended to accommodate the mutual interaction between matter and spacetime geometry.

\section{Extended and Backreacting Bodies}\label{REALISTIC}

In this section, I turn to the more general case of bodies whose stress–energy contributes to the spacetime geometry—that is, bodies which backreact. It is useful to distinguish two conceptually and mathematically distinct regimes in which such backreaction can be treated:

\begin{itemize}
    \item \textbf{Perturbative backreaction}: The metric is decomposed as $g_{ab}=g^0_{ab}+h_{ab}$, where $g^0_{ab}$ is a fixed background and $h_{ab}$ a small, \textit{linear} perturbation sourced by the body's own stress-energy. This regime captures \textit{gravitational self-interaction effects} without requiring the full non-linear EFEs to be solved. Although the body may be represented as a sharply localised source—often using a delta-function stress–energy tensor—such a representation is introduced only at the level of the linearised theory, where it can be justified as the limiting behaviour of an extended, smooth configuration. In fact, as shown by \cite{Gralla2008}, this point-particle description can be derived, not merely assumed, as a mathematically well-posed approximation to a compact body in a consistent perturbative framework. The resulting formalism avoids the conceptual failures that undermine geodesic motion as either an approximation or idealisation. It therefore provides the first step in a dynamically admissible hierarchy of approximations to natural motion. and occupies a physically meaningful intermediate position between test bodies and fully non-linear systems.

    \item \textbf{Non-perturbative backreaction}: In this regime, the body's stress–energy \textit{non-linearly }sources the spacetime geometry via the full EFEs. The geometry is entirely dynamical, and no background metric is specified a priori. Because of the non-linearity of the field equations, analytic solutions in this regime are rare. Nonetheless, important classes of exact solutions—including some cosmological models—fall into this category. But as will be shown, the matter sources' motion these models exhibit is not derived from realistic matter configurations: it is built into the model by assumption, not recovered from the underlying dynamics.

\end{itemize}

\subsection{Gravitational Self-Interaction in a Perturbed Background}\label{perturbed}

This subsection addresses the motion of small but extended bodies whose gravitational self-field is weak enough to be treated perturbatively. 
The central challenge is to describe how such bodies move under the influence of their own gravity without violating the dynamical structure of GR.

Two major formalisms approach this problem: the so-called \textit{MiSaTaQuWa formalism} and the related approach developed by Gralla and Wald. 
Both aim to describe the motion of compact, extended bodies under gravitational self-interaction. However, they differ crucially in how they treat the notion of a point particle. 
The MiSaTaQuWa formalism assumes from the outset a delta-function source, representing the body as point-like within the linearised Einstein equations. Although the referent is intended to be a compact extended body, the formalism \textit{postulates}, rather than derives, its representation as a structureless point-particle moving along a timelike worldline.

By contrast, the Gralla–Wald approach treats the body as extended throughout the derivation and obtains the point-particle description as a perturbative \textit{output}. 

\textit{At first order} in the perturbative expansion, the metric perturbation approximates the field generated by a point mass.

The worldline along which this field is supported is geodesic \textit{in the zeroth-order} limit, but is not a geodesic at higher orders due to self-force effects.
The point-particle representation thus emerges as a derived approximation of the perturbative regime, not as an exact representation of the body itself.

The distinction between assuming and deriving the point-particle representation is central, both mathematically and epistemologically, to understanding the validity and limitations of each formalism.

The MiSaTaQuWa formalism, developed by \cite{Mino1997} and independently by \cite{quinn1997axiomatic}, computes the motion of a small mass moving through curved spacetime under the influence of its own gravitational field. While the body is physically understood to be compact and extended, it is represented by a \textit{Dirac delta-function stress–energy} supported on a timelike worldline. This representation is introduced only within the linearised Einstein equations, where such distributional sources are mathematically well-defined. 
Since no delta function is inserted into the full non-linear field equations, the formalism does not violate the Geroch–Traschen theorem (\S\ref{gerochtraschen}).
However, the delta-function representation remains a modelling assumption, not a derived consequence. Nonetheless, the delta-function source remains a modelling assumption; the point-particle limit is not dynamically justified within \textit{full} GR, and the formalism is strictly off-shell.

Despite this limitation, MiSaTaQuWa successfully captures a key physical feature of self-interaction in curved spacetimes: the presence of \textit{ tail effects} \citep{Caldwell1993}. 
In curved spacetimes, field propagation violates Huygens’ principle and gravitational perturbations propagate not only on, but also inside, the light cone due to curvature-induced backscattering. Formally, this is a consequence of the fact that the Green's function of the wave operator has support inside the light cone.
As a result, the gravitational field can influence its own source at later times. These \textit{history-dependent contributions}, known as tail effects, mean that the body experiences delayed echoes of its own past gravitational field. First rigorously analysed in electrodynamics by \cite{DeWitt1960}, , and later extended to gravity in the MiSaTaQuWa framework, tail terms play an essential role in modelling dissipative dynamics, including gravitational-wave emission (e.g. in experiments like \href{https://lisa.nasa.gov}{LISA}).

In this setting, since the source is sharply localised being modelled by a delta-function stress-energy, its \textit{retarded} perturbation $h^{\rm tail}_{ab}$ diverges on the worldline, rendering the self-force ill-defined.\footnote{Just as the Coulomb field of a point charge diverges at the charge's location, so too does the linearised gravitational field become singular at the curve along which the delta source is concentrated. } This makes it impossible to directly substitute the raw perturbation into the equations of motion.

To address this, the retarded is decomposed into:
\begin{itemize}
    \item a \textbf{singular field} $h^{\rm tail, sing}_{ab}$, which carries the divergence but exerts no net force on the source;
    \item a \textbf{regular field} $h^{\rm tail,R}_{ab}$, which is smooth and governs the physical self-interaction.
\end{itemize}
This decomposition ensures causal consistency: the self-force at a given event is influenced only by past configurations of the source.

The resulting self-force, formalised in eq. (121) of \cite{Gralla2008}, is given by the \textit{MiSaTaQuWa equation of motion}:
\begin{equation}
    u^b\nabla_bu^a=F^a_{\rm self}=-\Big(g^{ab}+u^au^b\Big)\Big(\nabla^d h_{bc}^{\rm tail,R}-\frac{1}{2}\nabla_b h_{cd}^{\rm tail,R}\Big)u^c u^d,\label{misataquwamotion}
\end{equation}

which governs the leading-order deviation from geodesic behaviour of the body's motion, within the linearised theory. 

This equation of motion is coupled to a linearised Einstein equation (ibid., eq. 120), which governs the dynamics of the metric perturbation sourced by the body of mass $M$. Leaving aside the formal details, unnecessary for the discussion, the stress-energy source is represented schematically as Dirac delta distribution supported on the actual non-geodesic worldline $\xi$ of the body that \textit{includes} the self-force effects. Using a symbolic shorthand:
\begin{equation}
    T^{ab}\approx M\, {u^a \otimes u^b\delta^{(4)}_\xi}\label{misataquwasource}
\end{equation}
where with some abuse of notation $\delta_\xi^{(4)}$ denotes the four-dimensional Dirac delta distribution supported on $ \xi$;  $\tau$ denotes the proper time along $\xi$, and $u^a$ is tangent to $\xi$.\footnote{The energy-momentum tensor acts on an arbitrary smooth symmetric test tensor field $\Phi_{ab}$ as: $ T_{ab}[\Phi_{ab}] = M \int_\xi{\Phi_{ab}u^au^b\;d\tau}$. In coordinates: \begin{equation*}
    T^{\mu\nu}(x)=M\int{d\tau\;u^\mu(x,\tau)u^\nu(x,\tau)\cfrac{\delta^{(4)}(x^\mu-\xi^\mu(\tau)))}{\sqrt{-g}}}\quad \text{see eq. (2.19) in \cite{Mino1997}}
\end{equation*}}
The key point is that the worldline itself is not geodesic but incorporates first-order deviations due to self-interaction, making the system fully self-consistent to leading perturbative order (see below to a clarification of what self-consistent means in this context).

Importantly, from the perspective developed in this paper, it would be misleading to treat equation \eqref{misataquwamotion} as a \textit{correction} to an otherwise valid approximation. 
Geodesic motion, as argued in \S\ref{THEOREMS} and \S\ref{EXTENSION},  is not an approximation to be \lq{}corrected\rq{} but a formal artefact excluded by the full dynamics of GR. 
The MiSaTaQuWa equation does not correct geodesic motion; rather, it inaugurates a valid approximation framework for small, extended, backreacting bodies.
This makes it a consistent layer in the hierarchy of natural motion.

Yet the MiSaTaQuWa derivation rests on heuristic procedures which are open to criticism for being ad hoc and mathematically inconsistent. 
The delta-function source is postulated, not derived, and formal consistency is maintained by invoking the so-called \textit{Lorenz gauge relaxation}, which is the practice of using the Lorenz-gauged form of the linearised Einstein equation but not strictly imposing the Lorenz gauge condition. 
This approach is adopted to allow for non-geodesic motion, as a strict adherence to the full gauged linearised Einstein equation would otherwise enforce only geodesic paths. However, this technique raises concerns about the formal consistency of the derivation. 

These shortcomings are addressed by \cite{Gralla2008}, who construct a more rigorous framework for deriving the MiSaTaQuWa equation of motion.
These shortcomings are addressed in the Gralla–Wald formalism. 
Rather than introducing a delta-function source by assumption within the linearised theory—as in MiSaTaQuWa—they begin with a smooth stress–energy configurations $T^{(\lambda)}_{ab}$ in the full EFEs modelling a compact distribution of matter with mass and size both scaling with a parameter $\lambda$. 
They then show that, in the appropriate limit, the resulting linearised field equations are sourced by an effective delta-function localised on a timelike curve.

In particular, they consider  the limit of a one-parameter family of extended, smooth solutions $(\mathcal{M},g_{ab}(\lambda))$ sourced by $T^{(\lambda)}_{ab}$ .
For all $\lambda>0$, these are exact solutions of EFEs.
What they analyse is the limiting behaviour of this family of smooth metrics and stress–energies as $\lambda\to0$, in two complementary ways:

\begin{description}
    \item[Ordinary Limit:] As $\lambda\to0$, the body vanishes entirely, and  $g_{ab}(\lambda)\to{g}^{(0)}_{ab}$, a smooth vacuum background. In this limit, the worldline collapses to a geodesic of ${g}^{(0)}_{ab}$.
This derivation mimics that of Ehlers–Geroch  and suffers the same pathology: the limit recovers geodesic motion at zeroth order, and does so only by excising the material referent. Within my framework, this is a Type II failure of idealisation (limit system exists but lacks the limit property) and constitutes a pathological tracking approximation. 
In summary, the ordinary limit describes the overall behaviour of spacetime that ‘remains’ when the body disappears, allowing the basic geodesic motion to be derived within full GR.

    \item[Scaled Limit:] To overcome the inadequacy of the ordinary limit, Gralla and Wald define a scaled limit which permits the derivation of a valid first-order approximation for the motion of a small, compact bodies due to gravitational self-interaction.
As $\lambda\to0$,  the resulting geometry converges to a Schwarzschild solution with finite mass $M$. This shows that the shrinking body does not vanish, but becomes a small black hole in the limit—a coherent solution of the full EFEs.\footnote{The Schwarzschild geometry that emerges in the scaled limit of Gralla–Wald  is not imposed by hand, but emerges from the vanishingly small size of an extended body. In Einstein–Grommer proof, the singularity is a primitive: it is not derived from a limiting procedure on smooth bodies and does not lie within the manifold. The curve along which motion is attributed has no well-defined source, and thus no physical referent. This is what makes it conceptually problematic .This distinction is what allows the Gralla–Wald framework to justify the point-particle approximation dynamically and consistently—precisely what is missing in Einstein–Grommer case.} 
The point-particle approximation thus emerges not as a heuristic assumption but as a controlled limit of an extended, physically admissible configuration. 
Also, this makes it possible to define the physical properties of the body in a rigorous way—such as mass (via \citet{Arnowitt1960} or \citet{Komar1963} methods), spin, and higher multipole moments (if considered).\footnote{The motion of an extended body can be approximated to be governed by both its internal multipolar structure—arising from its finite size—and by the way its own stress–energy perturbs the geometry through which it moves.} These properties are crucial for establishing the body’s internal structure and for identifying the worldline that best represents its motion. 
In particular, Gralla and Wald choose a frame in which the mass dipole moment vanishes, which allows them to define a unique centre-of-mass worldline. The timelike worldline does not coincide with the background geodesic: its displacement encodes the first-order self-interaction effects.\footnote{These effects are especially significant in contexts like \textit{extreme mass-ratio inspirals} (EMRIs), where the body’s motion deviates from a background geodesic and such effect is measurable via the emission of gravitational waves \citep{Barack2009,AmaroSeoane2018}.}   

\end{description}
In summary, the scaled limit allows one to define the internal structure and physical parameters of the shrinking body.

When combined with the ordinary limit, the scaled limit yields an effective first-order linearised description in which the body behaves like a structureless point mass, whose equation of motion is eq.\eqref{misataquwamotion}. 
Gralla and Wald's analysis explicitly highlights that MiSaTaQuWa equations form an integro-differential system that self-consistently describes the body's motion and its own gravitational field. 
This \lq{}self-consistent\rq{} nature of the system \eqref{misataquwamotion}-\eqref{misataquwasource}, means that the linearised EFEs is a\textit{ nonlinear system} where the source itself depends on the solution.\footnote{This is not a contradiction. The term \lq{}linearised\rq{} describes the algebraic form of the differential operator acting on the metric perturbation. However, the "self-consistent" framework of the MiSaTaQuWa equations introduces a functional dependence of the source on the evolving solution, which makes the overall system of equations effectively non-linear, despite the linearszed appearance of the individual field equation.} 
In essence, eq. \eqref{misataquwamotion} describes a worldline that account for accumulated self-force effects.

Importantly, the delta-function stress–energy in eq.\eqref{misataquwasource} that sources the perturbation  is \textit{derived} as an emergent approximation, \textit{not assumed} as a fundamental input. This makes the framework dynamically consistent and free from the contradictions identified by the Geroch–Traschen theorem.
The self-force effects derived in this way belongs strictly to the perturbative regime of GR: it is meaningful only where a fixed background can be defined and deviations can be treated order by order. 
In the full non-linear theory, such an expansion is not possible, and the concept of self-force loses its validity. 

In summary, Gralla-Wald approach provides the mathematical setting in which the internal structure of the shrinking body can be preserved, physical properties can be defined, and gravitational self-force effects can be derived in a consistent and controlled way.

The intermediate regime presented in this subsection—perturbative self-interaction in a fixed background—marks a further erosion of the geodesic principle.  Unlike geodesic motion, which cannot be derived from any admissible dynamical model and lacks a valid target system, the MiSaTaQuWa equations, refined by Gralla and Wald, yield a physically meaningful approximation. They describe motion that is dynamically consistent with the Einstein equations, provided the body is sufficiently small and compact. As such, the motion of eq.\eqref{misataquwamotion} represents a valid layer within the hierarchy of natural motion developed in this paper.

\subsection{Full Backreaction: The Cosmological Case }\label{cosmology}

The case of cosmology provides a final and highly instructive test of the geodesic principle. Unlike the MiSaTaQuWa formalism, which models small, backreacting bodies in a perturbative regime, cosmological modelling typically concerns fully backreacting matter fields whose stress–energy determines the large-scale geometry. 
Here, the geodesic principle loses not only its justification, but even its coherence.
With no background connection to define curvature or parallel transport independently of the matter configuration, the very notion of geodesic motion becomes ill-defined.

This is most evident in the interpretation of the \emph{Hubble flow} in FLRW spacetimes, which is widely taken to represent a physically realised geodesic motion of cosmic matter.\footnote{The Hubble flow describes the large-scale motion of matter driven by the expansion of spacetime itself. More formally, it isolates the component of a body’s recessional velocity due to cosmic expansion, distinguishing it from peculiar motion arising from local gravitational interactions \citep{Ryden_2016}. }
On this reading,  the flow lines of pressureless dust in FLRW provide a textbook example of inertial motion: they are geodesics of a spacetime that solves the Einstein equations. The geodesic principle appears not only to survive, but to be dynamically vindicated.

However, this appearance is deceptive. The FLRW dust model does not arise as an approximation to realistic matter motion in the universe, nor does it represent a legitimate idealisation grounded in the dynamics of full GR.

To see this clearly, consider first what is actually being modelled.
The universe we observe is structured, clumpy, and anisotropic\textit{ at all observable scales}. 
Structure formation proceeded through the amplification of tiny initial \textit{perturbations}, giving rise to the cosmic web of filaments, clusters, and voids.
While the \textit{cosmological principle} posits that the universe is \textit{approximately} homogeneous and isotropic at large scales, the corresponding FLRW model is not dynamically derived from this inhomogeneous structure. It is imposed as a modelling assumption.

The question, then, is whether the geodesic motion of FLRW dust can be recovered as a valid approximation or idealisation of the actual matter flow. According to the framework adopted in this paper, geodesic motion qualifies as an approximation only if it describes the behaviour of a real target system—even if inexactly—and as an idealisation only if it emerges as an allowed limit property instantiated by an allowed limit system within the theory.

In both respects, the FLRW model fails.

In a series of work (\citealp{Buchert1997,Buchert2001,Buchert2020}) Buchert and collaborators show that the process of \lq{}averaging\rq{} an inhomogeneous matter distribution does not yield the FLRW dynamics, even in the large-scale limit.\footnote{The averaging procedure is a formal method designed to derive an \textit{effective dynamics} for an inhomogeneous universe by \textit{spatially averaging} its properties over specific domains. This approach aims to bridge the gap between complex, realistic inhomogeneous models and the simpler, homogeneous and isotropic FLRW models traditionally used in cosmology.}
Instead, new terms—called \emph{backreaction terms}—appear in the effective dynamics, reflecting the influence of shear, expansion rate fluctuations, and local curvature.\footnote{The authors also stress that even if these backreaction terms were assumed to be negligible or to cancel out for some reason, the averaged model is still different from the standard homogeneous-isotropic models because averaged energy and momentum conservation laws do not generally simplify to the homogeneous case for inhomogeneous fluids. For example, an averaged inhomogeneous radiation cosmos does not follow the evolution of the standard homogeneous-isotropic model.} 
These terms measure the departure from a standard FLRW cosmology, and they do not generically cancel.  
The assumption that such terms vanish on large scales, sometimes called the \textit{cosmological conspiracy}, while often adopted in practice, is \textit{not} dynamically derived from the theory but introduced as a modelling assumption, essentially presupposing the outcome rather than deriving it. 
In their words, it is a strong restriction of generality and not dynamically justified.
As Buchert and collaborators stress, the standard models are often presupposed rather than derived, and many simulations enforce FLRW behaviour by construction, rather than obtaining it as an emergent feature.

These results have direct epistemic consequences. The geodesic motion of FLRW dust does not approximate the behaviour of any real, inhomogeneous matter distribution. There is no physically admissible target system—no solution to the EFEs with realistic matter content—whose motion is well approximated by the FLRW dust flow. The model therefore fails as an approximation.
Nor can it be defended as an idealisation.  In Norton’s framework, this would require the geodesic property of FLRW dust to emerge as a well-defined limit of a family of solutions with increasing inhomogeneity resolution. But Buchert’s results indicate that such a limit fails to preserve the geodesic property: the terms that emerge from the averaging procedure do not vanish in the appropriate limit, and the effective motion deviates from geodesic flow. This constitutes a Type II failure in Norton’s taxonomy: no consistent limit system exists within GR that realises FLRW geodesic motion.
The FLRW dust geodesics are not the idealised motion of any real body or family of systems—they are formal artefacts, imposed rather than dynamically derived.

It might be objected at this point that the FLRW model is not merely a geometric construction, but a physical solution of GR. This is correct. The FLRW spacetime with pressureless dust is sourced by the stress–energy tensor in eq. \eqref{stressenergydust} When substituted into the full, non-linear Einstein equations, this yields the familiar Friedmann equations:

\begin{eqnarray}
\left( \frac{\dot{a}}{a} \right)^2 = \frac{8 \pi G}{3} \rho - \frac{kc^2}{a^2} + \frac{\Lambda c^2}{3}\\
\left(\frac{\ddot{a}}{a}\right) = -\frac{4 \pi G}{3} \left( \rho + \frac{3p}{c^2} \right) + \frac{\Lambda c^2}{3}
\end{eqnarray}

where $a(t)$ is the scale factor;  $t$ is the cosmological time; $\rho(t)$ and $p(t)$ are the density and pressure of the matter; $k=0,-1,+1$ is the spatial curvature parameter; $\Lambda$ is the cosmological constant \citep{weinberg1972gravitation}, and the constants have their usual meaning.\footnote{The value of $k$ does not fix the overall topology. In fact, different topological choices are possible for the same $k$: for example, a hyperplane (closed topology) is
characterised by curvature parameter $k= 0$, like a hyperplane (open topology).}. 
These equations govern the evolution of the spacetime geometry under the influence of a dust source. In this setting, as anticipated in \S\ref{gerochjang} the worldlines of the dust fluid are geodesics of the evolving metric by construction (see eq.\eqref{geodesicFLRW}). 
In this sense, the geodesic motion of the dust is not postulated, but dynamically derived from EFEs. At first glance, this may appear to undermine the argument developed here. 
But the geodesic flow of FLRW solution is a formal artefact of symmetry, not a feature of realistic backreaction. 
In generic backreacting systems, the absence of a fixed background geometry renders the geodesic equation ill-defined as a dynamical principle.

The issue is not whether geodesic motion can be derived in highly symmetrical models—it clearly can. 
The issue is whether such motion can be recovered from any realistic matter configuration through a process of approximation or idealisation. 
The FLRW solution assumes exact homogeneity and isotropy at all scales. It excludes shear, vorticity, anisotropic stresses, and local density contrasts. Its geodesic motion is valid only because the model \textit{assumes} away every feature that might disturb it. 
As shown in the Buchert work, also the averaged dynamics include non-zero backreaction terms, and the effective fluid flow is non-geodesic. 
The failure is therefore not internal to the FLRW solution itself, but concerns its relationship to the physical systems it is supposed to approximate or idealise. 
As a representation of actual matter motion in general relativity, FLRW geodesic flow has no referent.

This does not diminish the analytical utility of FLRW cosmology. It remains a powerful framework for parameter estimation, perturbative analysis, and the construction of large-scale structure surveys. But its geodesic motion must be understood in a different light. It is not a limiting behaviour of real matter, nor a coarse-grained approximation of inhomogeneous dynamics. In the terms developed throughout this paper, it is a formal construction—mathematically consistent, physically ungrounded.

\section{Natural Motion: A Layered Notion}\label{naturalmotion}

The preceding sections have shown that the geodesic principle—the idea that free bodies follow geodesics of a background metric—fails to represent a physically admissible notion of motion within GR. 
As we have seen, geodesic motion is neither an approximation (it lacks a target system) nor an idealisation (no consistent limit system exists).
What it offers is not a law of physical motion but a \textit{useful theoretical construct}—a formally elegant artefact that simplifies calculations, but lacks any admissible referent within the space of physically meaningful solutions to the EFEs.

What GR offers in its place is is not a \textit{single} principle, but a plurality of context-dependent motions: a \textit{hierarchy of representational regimes}, each valid under specific approximations about matter and geometry. 
This structure motivates a conceptual shift—from inertial motion to \textit{natural motion}—which I now articulate.

\begin{definition}
    \textbf{Natural Motion:} the motion of a material body as determined by the most appropriate approximation scheme for its physical properties, including internal structure, spatial extension, coupling to curvature, and self-interaction. Natural motion is not governed by a single dynamical law, but by a hierarchy of representational regimes, each valid within a specific physical domain.
\end{definition}

Rather than seeking a single universal equation of motion, we accept that what counts as a \lq{}natural trajectory\rq{} depends on the body in question and the physical regime in which it is represented.  A spinning test body, a test body with quadrupole structure, a perturbatively self-interacting object, and a fully backreacting configuration all demand distinct formalisms. Natural motion is thus intrinsically \textit{plural}. 
Importantly, my framework does not deny that there exists a \lq{}most fundamental\rq{} notion of natural motion—namely, the motion a body undergoes \textit{under gravity alone}, taking into account \textit{both} its internal structure and its dynamical coupling to spacetime. 
Actually, this is the most complete and physically significant regime. 
But even in this regime, there is no \textit{single} \lq{}master equation of motion\rq{} valid for all cases. Rather, each class of body—depending on its symmetry, spin, stress–energy distribution, or constitutive fields—requires its own distinct dynamical model. 
The plurality of natural motion is thus not a by-product of approximation alone; it is a structural feature of general relativistic dynamics, rooted in the diversity of admissible matter types. It reflects a deeper \textit{ontological plurality}.

Assuming a given class of bodies, each layer in the hierarchy of approximations corresponds to a physically consistent regime within GR. \textit{These are not successive corrections} to geodesic motion, but \textit{independent dynamical frameworks} in their own right. It is worth repeating what I stated in \S\ref{REALISTIC}: geodesic motion cannot serve as the base of a dynamical expansion. The language of \lq{}corrections\rq{} obscures the fact that geodesic motion lies \textit{outside} the physically meaningful hierarchy of approximations.

What unifies these regimes is not a common mathematical structure, but a shared epistemic status: each describes the motion of bodies subject only to gravity, in a way that remains consistent with the dynamical constraints of the Einstein equations. Natural motion is defined by \textit{physical admissibility}, not by geometric simplicity.

The shift from inertial to natural motion thus marks a change not only in vocabulary, but in how motion is conceptually and physically framed. 

In the standard relativistic picture, inertial motion is defined by geodesic trajectories and relative to local INRFs. But since no physically real or even idealised body can follow such a trajectory, the referential structure collapses. 
But as shown in \S\ref{geodesic} and \S\ref{EQUIVALENCE}, such frames refer to  no possible measurement apparatus. They are mathematically well defined as a description of vanishing connection coefficients, but physically vacuous. As such, the local INRF must be understood as a \textit{purely formal construct}, used to encode information about local geometric structure, not about the motion of material systems. It serves as a formal tool, not a physical frame.
Again, this does not imply that the concept of a local inertial frame is ill-formed, only that it lacks any physical realisation.

Natural motion, by contrast, requires \textit{no privileged frame or trajectory}, but a layered hierarchy of physically meaningful regimes, each describing the motion of bodies under gravity alone. 
As already stated in \S\ref{deviation}, one may still introduce local INRF structures to aid interpretation or calculation, as shown in the geodesic deviation formalism. 
However, this introduction is a matter of methodological convenience, not an physical statement.

This conceptual shift culminates in a new foundational principle, one that supersedes the Principle of Inertia as defined in the relativistic framework via definitions PIN (v.4)-(v.5) (\eqref{PIN4}-\eqref{PIN5}):

\begin{definition}
\textbf{Principle of Natural Motion (PNM) (v.1):} A body maintains natural motion if and only its motion is determined by no interaction other than gravity. The notion of natural motion is not unique, but varies across body types and levels of approximation.

\end{definition}
\begin{definition}
\textbf{PNM (v.2):} Natural motion is not defined relative  to any privileged class of frames, but relative to a physically justified approximation regime. References to geodesic motion within such formal schemes play only counterfactual role: they provide formal scaffolding, not physically meaningful standards of motion.
\end{definition}

Unlike PIN (v.4)-(v.5), PNM (v.1)-(v,2) is not anchored in any preferred trajectory type, nor does it presuppose the existence of local INRFs. 
It expresses a different kind of commitment: that the motion of bodies \textit{under gravity} is to be described by approximation regimes consistent with EFEs, and that no such regime yields geodesic motion as a valid limit.\footnote{This also suggest that the unification that GR achieves is not one of gravity and inertia, as traditionally understood (see \citealp{LEHMKUHL2014}), but of gravity and natural motion—a concept with physical content, layered structure, and dynamic validity. This will be the focus of future work.}

This principle does not identify a \textit{new} kind of motion. It describes the \textit{only} physically meaningful kind of motion permitted by GR.  This is not merely a conceptual rebranding. It reflects a deeper epistemic transformation.
What was once thought to be the purest expression of natural motion—the geodesic trajectory—is now  revealed as physically vacuous, a \textit{formal artefact}.
What replaces it is not a \textit{new}, better, universally valid trajectory, but a layered structure of regimes each tailored to describe the motion of systems within a precise domain of validity. 
\textit{For any given class of body}, the most accurate expression of its natural motion arises from the regime that includes both its internal structure and its backreaction on the metric.

As such, natural motion replaces the search for a universal trajectory law with a pluralistic model of dynamical representation.
This reflects a deeper epistemic stance: motion in GR is not anchored in a formal, \textit{a priori preferred} construction, but \textit{derived} from the dybamical structure of the theory.
The PNM inherits the role that the geodesic principle and the PIN aspired to play—while jettisoning the formalist assumptions that made that aspiration untenable.

The layered hierarchy of motion in GR can now be schematically summarised in Table \ref{tab:regimes_GR_summary}.
The upshot is clear. Natural motion is not a refinement of the geodesic principle. It is its systematic displacement.

\begin{table}[ht]
\centering
\renewcommand{\arraystretch}{1.4}
\begin{tabular}{|>{\raggedright\arraybackslash}p{4.8cm}|>{\raggedright\arraybackslash}p{4.5cm}|>{\raggedright\arraybackslash}p{6.5cm}|}
\hline
\textbf{Body Type} & \textbf{Applicable Framework} & \textbf{Key Features of Motion} \\
\hline
\multicolumn{3}{|c|}{\textbf{--- Not Part of the Natural Motion Hierarchy ---}} \\
\hline
Point-like, non-backreacting, spinless & Geodesic equation & Formal construct only. No real or idealised body moves in this way. Not a valid approximation or idealisation.\\
\hline
\multicolumn{3}{|c|}{\textbf{--- Natural Motion Regimes ---}} \\
\hline
Extended, non-backreacting, spinning& MPD equations& Spin–curvature coupling $\&$ Quadrupole–curvature coupling; fixed background.\\
\hline
Extended, non-backreacting & Geodesic deviation& Internal tidal effects; fixed background.\\
\hline
Extended, backreacting (perturbative)& MiSaTaQuWa formalism & Self-interaction: tail terms from perturbation field\\
\hline
Extended, backreacting (non-perturbative)& Non-linear, general-relativistic models, often simplified in Cosmological cases& Full coupling to spacetime geometry via the EFEs\\
\hline
\end{tabular}
\caption{Hierarchy of physically admissible regimes of motion in GR. Geodesic motion is shown above the horizontal break to emphasise its exclusion from the hierarchy of natural motion: it is neither an approximation nor an idealisation. Below the break, natural motion emerges in layered regimes of consistent dynamical representations, each tied to specific physical assumptions about body'sstructure and backreaction.}
\label{tab:regimes_GR_summary}
\end{table}

\section{Conclusion}

My inquiry ends here, provisionally. On the 2,000-year road from Aristotle to Einstein, I have revisited familiar landmarks and begun to chart a less-travelled route: one in which inertial motion and natural motion—long treated as coextensive—are systematically disentangled. What began as a foundational tenet of mechanics, the Principle of Inertia, emerges in the context of general relativity as a formal artefact: mathematically elegant, but physically unrealised.

The investigation began in \S\ref{classicalinertia}, which traced the historical and conceptual evolution of the Principle of Inertia from its classical roots to its pre-relativistic reformulations, revealing deep instabilities in both its content and its foundational role.
I showed that traditional formulations of inertial motion---via uniform motion, absence of force, or privileged reference frames---suffer from fatal circularities.
\S\ref{lawvsprin} clarified inertia’s epistemic status—specifically, whether it should be understood as an empirical law or as a structural principle. 
Two major interpretive approaches were introduced: the \textit{law-like} and the \textit{principle-like}. 
Under the law-like reading, inertia is treated as a descriptive regularity—either in terms of unforced motion (\textit{law-based} approach) or as the consequence of a fixed geometric background (\textit{structure-based} approach). Both variants seek to define inertial frames either empirically or geometrically, but ultimately fail to offer a non-circular characterisation. 
In contrast, the principle-like reading reconceives inertia as a constitutive feature of the theory, encoded in its dynamical symmetries. 
I proposed that Jacobs’ \textit{symmetry-based} approach naturally supports this principle-like understanding of inertia.

\S\ref{geodesic}, examined the relativistic generalisation of inertia through the \textit{geodesic principle}. While this principle avoids the circularities of classical definitions by identifying inertial motion with geodesic motion of the Levi–Civita connection, it collapses into \textit{triviality}: geodesic motion merely restates the structure of the connection, offering no genuine explanation of how bodies move. This triviality became especially evident in the formal definitions I introduced, such as PIN (v.4) and PIN (v.5), which recast inertial motion either as the absence of non-gravitational couplings in the Lagrangian or as geodesic motion relative to a local inertial frame.
The reliance on \textit{local inertial frames}, which is supported by the Equivalence Principle, further complicates the picture. 
As \S\ref{EQUIVALENCE} argued, the very notion of locality invoked—whether pointwise or along a geodesic—is physically impractical. It licences only mathematical constructs, not empirically realisable systems.

\S\ref{THEOREMS} clarified the central philosophical thesis of the paper: that geodesic motion in GR is neither an approximation to, nor an idealisation of, the motion of real bodies.
Drawing on Tamir’s distinction between limit proofs and singularity proofs, and on Norton’s framework for diagnosing failures of idealisation and approximation, I examined four canonical derivations of the geodesic principle: Geroch–Jang, Ehlers–Geroch, Einstein–Grommer, and Geroch–Traschen. Each of these strategies was shown to fail not only as a derivation of the geodesic principle—Tamir’s original claim—but also as a justification for treating geodesic motion as a valid idealisation or approximation within GR.
 
The Geroch–Jang theorem presupposes the test-body regime by inserting non-zero stress–energy into a fixed background without any dynamical justification, violating the Einstein equations and producing what I termed an \textit{off-shell failure} of approximation (\S\ref{gerochjang}). 

The Ehlers–Geroch theorem introduces a converging sequence of spacetimes, but in the limit the stress–energy vanishes: the limit system is a vacuum spacetime that contains no body at all and therefore cannot bear the geodesic property—an instance of \textit{Norton’s Type II failure} of idealisation (\S\ref{ehlersgeroch}).

The Einstein–Grommer strategy removes the body from the manifold altogether and attributes geodesic motion to a curve lying outside the spacetime. This is a paradigmatic case of \textit{pathological tracking}: the motion of the body is not approximated but erased (\S\ref{einsteingrommer}).

Finally, the Geroch–Traschen theorem proves that no distributional stress–energy supported on a curve can satisfy the field equations: the limit system needed to realise geodesic motion simply does not exist—an instance of \textit{Norton’s Type I failure} (\S\ref{gerochtraschen}). 

Across all four cases, geodesic motion emerges as a well-defined mathematical construct that is neither instantiated by any physically admissible system nor borne by any coherent idealisation. Its physical bite is thus illusory.

In \S\ref{EXTENSION} I examined the motion of spatially extended test bodies in GR, whose internal structure induces systematic departures from geodesic motion even in the absence of backreaction. These systems occupy an intermediate regime: they are not point-like, but they do not source curvature. 
The section analysed two complementary frameworks that yield the first physically meaningful approximations to natural motion in curved spacetime. 
The \textit{Mathisson–Papapetrou–Dixon (MPD) formalism} captures how internal spin and higher multipole moments couple to curvature, modifying the motion of the centre of mass (\S\ref{MPD}). 
The \textit{geodesic deviation formalism}, by contrast, models how tidal effects—arising from curvature gradients—induce relative acceleration across the body's interior modelled as a congruence of geodesics. 
Moreover, by analysing the role of \textit{torsion}, I showed that even the geometry used to define free fall expose the geodesic principle as inadequate: the very meaning of natural motion depends on the structural features of the spacetime model (\S\ref{torsion}). 
These findings mark a turning point in the paper: they reveal that geodesic motion lies entirely outside the approximation hierarchy grounded in physically admissible systems.

\S\ref{REALISTIC}, completed the construction of the layered natural motion by addressing the motion of extended, backreacting bodies in GR.
I distinguished two conceptually distinct regimes of backreaction. 
The first—\textit{perturbative backreaction}—describes small, compact bodies whose self-gravity can be modelled as a linear perturbation of a fixed background (\S\ref{perturbed}). 
I examined the \textit{Mino-Sasaki-Tanaka-Quinn-Wald (MiSaTaQuWa) formalism}, which introduces a delta-function stress–energy to compute self-force effects, but does so only within the linearised Einstein equations. 
While this approach captures essential features of gravitational self-interaction—such as history-dependent \textit{tail terms}—its use of a point-particle source is assumed rather than derived. 
This gap is closed by the \textit{Gralla–Wald construction}, which rigorously derives the MiSaTaQuWa formalism as the limit of a one-parameter family of smooth, extended, on-shell solutions. 
This marks a pivotal moment in the argument: unlike geodesic motion, the MiSaTaQuWa worldline represents a real, backreacting system within GR and thus occupies a legitimate position within the hierarchy of natural motion.

By contrast, the second regime—\textit{non-perturbative backreaction}—reveals the illusory character of geodesic motion in the standard cosmological FLRW model (\S\ref{cosmology}). 
I showed that the geodesic Hubble flow of cosmic dust, though often cited as the canonical example of inertial motion in GR, does not approximate any realistic inhomogeneous matter configuration. Averaging procedures over inhomogeneous spacetimes generically yield non-geodesic effective flows, as shown by Buchert and collaborators. This amounts to a Type II failure of idealisation: even in the large-scale limit, no consistent system realises the geodesic property. The FLRW geodesics are thus revealed to be symmetry-induced artefacts with no physical referent—neither real nor idealised. 
Nevertheless, the fully non-linear regime remains central to the overall framework. Far from being excluded, fully non-perturbative, backreacting systems—despite their analytic intractability—constitute the most general and fundamental level of natural motion.

Across both regimes, this section confirmed that geodesic motion cannot be salvaged as a model of free fall in backreacting systems. They complete the conceptual arc begun in earlier sections: the rejection of geodesic motion is not a renunciation of dynamical realism, but a pathway to recovering it on stronger grounds.

\S\ref{naturalmotion} completed the constructive arc of the paper by explicitly articulating the layered concept of \textit{natural motion} that had been progressively developed across \S\S\ref{EXTENSION}-\ref{REALISTIC}.
Drawing together the distinct approximation regimes explored earlier—ranging from structured test bodies to perturbative and fully nonlinear backreaction—this section defined natural motion as the \textit{system-specific}, dynamically consistent motion of a body under gravity alone.  
It was shown that no single law or trajectory captures this plurality. Instead, each level of approximation corresponds to a distinct formalism that respects both the matter configuration and the constraints of the EFEs.
Importantly, this layered pluralism is not merely approximation-theoretic in character. Even in the most complete and physically fundamental regime—namely, the fully non-linear dynamics of backreacting bodies under gravity alone—there exists no single, universal equation of motion. This reveals a deeper \textit{ontological plurality}: the diversity of natural motion reflects not just the limitations of approximation, but the structural richness of general relativistic dynamics itself.
This plural structure was formalised in a new \textit{Principle of Natural Motion}, which does not rely on privileged trajectories or frames and displaces the geodesic principle as the foundational principle of free motion in GR.  In this framework, geodesic motion is no longer the core ideal to be corrected or recovered; it is an artefact excluded by every admissible regime. Natural motion thus replaces inertial motion as the physically meaningful standard for describing how bodies move under gravity in GR.

What, then, becomes of inertia? The lesson that GR seems to teach us is that
 the concept persists, but only as a formal by-product of GR’s geometric structure. It serves as a calculational aid, not a physically instantiated principle. The rhetorical force of the geodesic principle masks its physical vacuity.
What GR offers instead is a richer picture: one that replaces the inertial framework with a plurality of natural motions, each dynamically valid within its regime of approximation.

This conclusion is not a terminus. The distinctions drawn here open several paths for future research.

One concerns the extension of the natural motion framework to field theory. What does it mean for a field configuration—governed by Euler–Lagrange dynamics—to evolve \textit{naturally}, rather than \textit{inertially}? Can the distinctions between the various layers of natural motion be framed for field degrees of freedom? 
Such questions invite deeper inquiry into the dynamics of fields in curved spacetime.

A second direction concerns generalisation beyond GR. I already mentioned the role of torsion. 
In theories such as Newton–Cartan gravity or even in equivalent formulations of GR \citep{BeltrnJimnez2019,Wolf2024}, one may ask whether similar tensions arise between inertial and natural motion.  The framework developed here could offer new criteria for distinguishing between formal constructs and dynamically meaningful trajectories in alternative gravitational theories.

Finally, this work suggests a broader interpretive shift. The unification that GR achieves is not between gravity and inertia—as traditionally claimed—but between gravity and natural motion.
The true lesson is that motion under gravity must be represented not by a privileged class of curves, but by a layered system of approximation regimes: structurally rich, dynamically consistent, and physically grounded.

\clearpage
\bibliography{BIB2.bib}

\end{document}